\documentclass[11pt,a4paper]{article}
\usepackage[margin=2.5cm]{geometry}
\usepackage[T1]{fontenc}
\usepackage[english]{babel}
\usepackage{times}

\usepackage{amsmath,amsthm,amssymb} 
\usepackage{stmaryrd} 
\usepackage{dsfont}
\usepackage{enumitem}
\usepackage{colortbl} 
\usepackage{graphicx}
\usepackage{multicol,multirow}

\usepackage[hyperfootnotes=false,breaklinks]{hyperref}
\hypersetup{
    colorlinks,%
    citecolor=blue,%
    filecolor=blue,%
    linkcolor=blue,%
    urlcolor=blue
}

\newcommand{\D}{\displaystyle}
\newcommand{\X}{\mathbf{X}}

\makeatletter
\renewcommand*{\@fnsymbol}[1]{\ensuremath{\ifcase#1\or 1\or 2\or 3\or *\or \dagger\else \@arabic{#1}\fi}}
\makeatother

\title{Fiber crosslinking drives the emergence of order in a 3D dynamical network model}

\author{Pauline Chassonnery\thanks{RESTORE, Universit\'e de Toulouse, Inserm U1031, EFS, INP-ENVT, UPS, CNRS ERL5311, Toulouse, France.}$^{\:\: ,}$\thanks{Inria Paris, team MAMBA, Sorbonne Universit\'e, CNRS, Universit\'e de Paris, Laboratoire Jacques-Louis Lions UMR7598, F-75005 Paris.} \and Jenny Paupert\footnotemark[1] \and Anne Lorsignol\footnotemark[1] \and Child\'erick S\'everac\footnotemark[1] \and Marielle Ousset\footnotemark[1] \and Pierre Degond\thanks{Institut de Math\'ematiques de Toulouse ; UMR5219, Universit\'e de Toulouse ; CNRS, UPS, F-31062 Toulouse Cedex 9, France.} \and Louis Casteilla\footnotemark[1]$^{\:\: ,}$\thanks{co-last authors} \and Diane Peurichard\footnotemark[2]$^{\:\: ,}$\footnotemark[4]$^{\:\: ,}$\thanks{corresponding author : diane.a.peurichard@inria.fr} \\
}

\date{}

\begin{document}
\renewcommand\labelitemi{$\bullet$}

\maketitle

\begin{abstract}
The Extra-Cellular-Matrix (ECM) is a complex interconnected $3$D network that provides structural support for the cells and tissues and defines organ architecture key for their healthy functioning. However, the intimate mechanisms by which ECM acquire their $3$D architecture are still largely unknown. In this paper, we address this question by means of a $3$D individual based model of interacting fibers able to spontaneously crosslink or unlink to each other and align at the crosslinks. We show that such systems are able to spontaneously generate different types of architectures. We provide a thorough analysis of the emerging structures by an exhaustive parametric analysis and the use of appropriate visualization tools and quantifiers in $3$D. The most striking result is that the emergence of ordered structures can be fully explained by a single emerging variable : the proportion of crosslinks in the network. This simple variable becomes an important putative target to control and predict the structuring of biological tissues, to suggest possible new therapeutic strategies to restore tissue functions after disruption, and to help in the development of collagen-based scaffolds for tissue engineering. Moreover, the model reveals that the emergence of architecture is a spatially homogeneous process following a unique evolutionary path, and highlights the essential role of dynamical crosslinking in tissue structuring.
\end{abstract}

\noindent{\makebox[1in]\hrulefill}\newline
\newline\textit{Subjects} Biophysics, Biomechanics, Computational Biology.

\section{Introduction}

The adequate architecture of any organ is mandatory for their efficient physiological function. It depends mainly on the structure of the extracellular matrix (ECM) which provides spatial information for cells and largely participate to mechanical constraints~\cite{Theocaris2016}. For example, in epithelial tissues, a basket-weave structure in the skin plays an important role in preserving the barrier function of human skin~\cite{Goto2020}, while alignment as well as accumulation of ECM observed in fibrosis lead to a loss of function~\cite{Mays1991,Piotrowski2017}. The ECM is a dynamical three-dimensional network consisting of interacting extracellular macromolecules such as collagen, enzymes and glycoproteins that provide structural and biochemical support to surrounding cells~\cite{Mays1991}. ECM fibers are interconnected by molecular bonds, i.e. crosslinks, that confer connectivity and elasticity throughout the ECM network. 

This network structure is in constant remodeling, which is crucial to maintain tissue integrity and function. Crosslinks, however, can unbind spontaneously or under tension, which leads to viscoplastic material responses, such as softening and tension relaxation~\cite{Malandrino2019}. Fibrosis and aging are also characterized by an increase of enzymatic and non-enzymatic crosslinks~\cite{Kar2021,Diller2022} and this increase in crosslinking prevents ECM degradation by matrix metalloproteinases, both events leading to a decrease of ECM remodeling~\cite{Levantal2009}. Altogether, these events induce greater stiffness and the arrangement of the collagen fibers becomes less organized and more loose and fragmented, hence weakening tissue integrity and strength~\cite{Wilson2013,Bailey1999}. An understanding of the fundamental organizing principles of ECM structure in three dimensions also helps in apprehending the complex dynamics of pathological tissues from degenerative diseases or tumor~\cite{Levantal2009}.

Because the global architecture of fiber networks seems to be fundamental for controlling tissue functions, modeling the process of ECM structure emergence will greatly improve our understanding of tissue biology and plasticity in physiological or pathological conditions. Numerous models of fiber networks can be found in the literature. Due to their simplicity and flexibility, the most widely used models are Individual Based Models (IBM), which describe the behavior of each agent (e.g. a fiber element) and its interactions with the surrounding agents over time~\cite{Drasdo2003,Hwa2009}. However, IBM have a high computational cost which can become intractable when studying large scales, either spatial or temporal or for systems composed of too many agents. In such cases, continuous or mean-field kinetic models may be preferred~\cite{Dal1999,Did2006,Riz2012,Unt2014,Barre2016} since they are less costly, but at the expense of a loss of information at the individual level. Since it is well acknowledged that microstructure configurations modulate the macroscopic properties of crosslinked fiber networks~\cite{Lie2010}, preserving the microscopic level description is of great importance to model tissue emergence.

Most of the computational models developed thus far for mimicking ECM networks are two-dimensional~\cite{Hea2003,Wil2003,Did2006,Heuss2006,Bai2011,Sha2012,Ber2013,Boi2013,Alo2014,Unt2014,Diane2017}. Few studies have been conducted on $3$D models~\cite{Sty2007,Huis2007,Ast2008,Bro2012,Har2013,Lee2014,Lin2015,Picu2018}, although these are expected to yield different, more realistic results than $2$D ones since they better mimic biological structures themselves immersed in $3$D environments. One of the reasons for fewer $3$D models is the great increase in the number of agents needed to achieve a given spatial density and thus in the associated computational cost. Another reason is the lack of high quality data on ECM organization in $3$D. However, the latter is becoming less and less of an issue with recent improvements in high resolution $3$D imaging and its availability. Among existing $3$D models, few of them feature dynamical crosslinking of ECM components. In~\cite{Ast2008,Har2013,Mak2020}, various models of $3$D fibrous networks composed of permanent or transient crosslinks (remodeling) are proposed. However, most of these models feature ECM remodeling in reaction to external factors (applied load~\cite{Ast2008}, migrating cells~\cite{Har2013}, contractile cells~\cite{Mak2020}), and the literature so far provides little cues on the mechanisms underlying fiber self-organization. 

In the present paper, we test the hypothesis that fiber macrostructures could spontaneously emerge without appealing to contact guidance or external mechanical challenges, as a result of simple mechanical interactions between the fiber elements composing the ECM network. We assess this hypothesis by means of a three dimensional model that is a $3$D extension of the two-dimensional model of ECM presented in~\cite{Diane2017} in the frame of adipose tissue morphogenesis, which was validated against biological data. ECM fibers are discretized into unit fiber elements, consisting of non-stretching and nonflexible sphero-cylinders with the ability to spontaneously link to and unlink from their close neighbors. This dynamical crosslinking mechanism allows us to model both the overall temporal plasticity of the network and the complex physical properties of biological fibers such as elongation, bending, branching and growth, thus compensating our minimalistic description of the fiber units.

Through computational simulations and exhaustive parametric analysis, we demonstrate that organized macrostructures can spontaneously emerge without external guidance. This study provides a comprehensive view on the role of ECM connectivity on tissue architecture emergence. The model first reveals that dynamic remodelling is essential for the generation of ordered ECM structures. Moreover, we surprisingly find that for dynamical networks, tissue architecture at equilibrium is simply controlled by the proportion of crosslinks in the network, independently of the amount of fibers or the remodelling speed of the network. These major results show that the emergence of ordered structures in biological fiber networks could be principally driven by the proportion of crosslinks they contain. This simple emerging variable therefore becomes an important putative target to control and predict the development of the architecture of biological tissues. Because of its simplicity, this variable is amenable for experimental measurements and could represent a major target for the development of therapeutic drugs enabling to induce tissue recovery after injury, prevent tissue degradation during ageing, or help in the design of engineering collagen scaffolds for tissue regeneration. With this in mind, we perform a deep exploration of the model parameters and use quantitative tools to characterize as precisely as possible how the different spatial structures emerge as function of the intrinsic parameters of our networks.

\section{Models and methods}

\subsection{Description of the model}

We model the complex fiber structures by $\textrm{N}_{\textrm{fib}}$ unit fiber elements consisting of line segments of fixed and uniform length $L_{\textrm{fib}}$. They are represented by the position of their centers $\X_k(t) \in \mathbb{R}^3$ and their directional unit vectors $\omega_k(t)$. These elements are non-oriented, meaning that we can restrict the phase-space of the directional vectors to the half unit sphere $\mathbb{S}_{2}^{+}$.

We include the following biological features~: \textit{(i) Fiber resistance to pressure} is modelled by a short-range repulsive force between pairs of fibers. Indeed, from a physical point of view, each fiber occupies a given volume from which other fibers are excluded. However, since implementing a strict sterical exclusion constraint would have a high computational cost, we settled for a repulsive interaction allowing some interpenetration between fibers. We assume that the intensity of the force field generated by a fiber decreases linearly with the distance to this fiber, thus displaying sphero-cylindrical isolines. We denote by $\alpha_{\textrm{rep}}$ the maximal intensity of this force field, which is reached on the fiber, and by $R_{\textrm{fib}}$ the threshold beyond which the force field vanishes (this can be regarded as the ``width'' of the fiber).

Given two fibers $k$ and $m$, we denote by $\X_{k,m}$ the point of fiber $k$ closest to fiber $m$ (see annex~\ref{annex_closestpoints} for the actual computation of this point). If $||\X_{k,m}-\X_{m,k}|| \le 2 R_{\textrm{fib}}$, fiber $k$ sustains from fiber $m$ a repulsive force~:
\begin{equation}
    \mathbf{F_{k,m}^{\textrm{rep}}} = \alpha_{\textrm{rep}} \left(2 R_{\textrm{fib}} - ||\X_{k,m} - \X_{m,k}||\right)^{3/2} \sqrt{2 R_{\textrm{fib}}} \times \frac{\X_{k,m} - \X_{m,k}}{||\X_{k,m} - \X_{m,k}||},
\end{equation}
which is applied at point $\X_{k,m}$, thus inducing a rotational torque~:
\begin{equation}
    \mathbf{T_{k,m}^{\textrm{rep}}} = \left(\X_{k,m} - \X_k\right) \wedge \mathbf{F_{k,m}^{\textrm{rep}}},
\end{equation}
on fiber $k$.

\textit{(ii) Fiber growth, elongation and ability to bend} are modelled by allowing two fibers closer than a certain threshold $d_{\textrm{link}}^{\max}$ to create a crosslink. A crosslink is defined as a spring of stiffness $\alpha_{\textrm{rest}}$ and unloaded length $d_{\textrm{link}}^{\textrm{eq}}$, fixed to the two points of the crosslinked fibers that were closest at the time of its creation. For two linked fibers $k$ and $m$, let us denote by $\X_{k,m}^l$ the point of fiber $k$ that was closest to fiber $m$ at the time of the link creation~: the elastic restoring force sustained by fiber $k$ due to its link with fiber $m$ is equal to
\begin{equation}
    \mathbf{F_{k,m}^{\textrm{rest}}} = \alpha_{\textrm{rest}} \left(d_{\textrm{link}}^{\textrm{eq}} - ||\X_{k,m}^l - \X_{m,k}^l||\right) \frac{\X_{k,m}^l - \X_{m,k}^l}{||\X_{k,m}^l - \X_{m,k}^l||},
\end{equation}
and induces a rotational torque on fiber $k$~:
\begin{equation}
    \mathbf{T_{k,m}^{\textrm{rest}}} = \left(\X_{k,m}^l - \X_k\right) \wedge \mathbf{F_{k,m}^{\textrm{rest}}}.
\end{equation}

To ensure coherence between the different features of the model, we require that $2 R_{\textrm{fib}} \le d_{\textrm{link}}^{\textrm{eq}} \le d_{\textrm{link}}^{\max}$. Linking and unlinking processes follow Poisson processes with frequencies $\nu_{\textrm{link}}$ and $\nu_{\textrm{unlink}}$ respectively. As a result, the linked fiber ratio $\D \chi_{\textrm{link}} = \frac{\nu_{\textrm{link}}}{\nu_{\textrm{link}} + \nu_{\textrm{unlink}}}$ represents the equilibrium fraction of linked fibers among the pairs of neighbouring fibers. Several consecutively crosslinked fiber units would model a long fiber having the ability to bend or even take possible tortuous geometries. As the number of crosslinks attached to a given fiber is not limited, we can also account for fiber branching. Therefore, the crosslinking process will model fiber elongation~\cite{Cia2009} and symmetrically, spontaneous unlinking of pairs of crosslinked fibers will allow for fiber breakage describing ECM remodeling processes.

\textit{(iii) Fiber stiffness}, that is the ability of biological fibers to offer a certain resistance to bending, is accounted for by subjecting pairs of linked fibers to a torque at their junction. This torque vanishes when the fibers are parallel, and consequently acts as a linked-fiber alignment mechanism. It is characterized by a stiffness parameter $\alpha_{\textrm{align}} > 0$ playing the role of a flexural modulus~: the larger $\alpha_{\textrm{align}}$, the more rigid the fiber network. Given two linked fibers $k$ and $m$, the torque sustained by the fiber $k$ is such that, $\forall \mathbf{u} \in \mathbb{R}^3$,
\begin{equation}
    \mathbf{T_{k,m}^{\textrm{align}}} \wedge \mathbf{u} = \alpha_{\textrm{align}} \left( (\omega_k \wedge \widetilde{\omega}_m) \wedge \mathbf{u} + \frac{1 - |\omega_k \cdot \omega_m|}{||\omega_k \wedge \omega_m||^2} (\omega_k \wedge \widetilde{\omega}_m) \wedge ((\omega_k \wedge \widetilde{\omega}_m) \wedge \mathbf{u}) \right),
\end{equation}
where $\widetilde{\omega}_m = \text{sign}(\omega_k \cdot \omega_m) \cdot \omega_m$ so that there is no preferential orientation.

\textit{(iv) Over-damped regime} (i.e. negligible acceleration during one time-step) is a physically reasonable hypothesis in the case of biological fibers, since they are surrounded by a rather thick medium which induces high friction. We assume that the friction sustained by an infinitesimal element of a fiber follows a Stokes law with friction coefficient $\mu_{\textrm{fib}}$. The total friction force sustained by a fiber $k$, computed by integrating this law on the whole length of the fiber, is equal to~:
\begin{equation}
    \mathbf{F_k^{\textrm{fric}}} = - \mu_{\textrm{fib}} L_{\textrm{fib}} \frac{d\X_k}{dt}
\end{equation}
and the associated rotational torque is equal to~:
\begin{equation}
    \mathbf{T_k^{\textrm{fric}}} = - \mu_{\textrm{fib}} L_{\textrm{fib}}^3 \omega_k \wedge \frac{d\omega_k}{dt}.
\end{equation}

We denote by $p_{k,m}(t)$ the linking state of fibers $k$ and $m$, that is $p_{k,m}(t)$ is equal to $1$ if fibers $k$ and $m$ are linked at time $t$ and to $0$ otherwise. The fundamental principle of dynamics in the over-damped regime gives rise to the following set of differential equations~:
\begin{equation}
	\left\lbrace\begin{array}{l}
		\D \mu_{\textrm{fib}} L_{\textrm{fib}} \frac{d\X_k}{dt}(t) = \sum_{m=1}^{\textrm{N}_{\textrm{fib}}} \left( \mathbf{F_{k,m}^{\textrm{rep}}}(t) + p_{k,m}(t) \mathbf{F_{k,m}^{\textrm{rest}}}(t) \right) \\
		\D \mu_{\textrm{fib}} L_{\textrm{fib}}^3 \frac{d\omega_k}{dt}(t) = \sum_{m=1}^{\textrm{N}_{\textrm{fib}}} \left( \mathbf{T_{k,m}^{\textrm{rep}}}(t) + p_{k,m}(t) \left(\mathbf{T_{k,m}^{\textrm{rest}}}(t) + \mathbf{T_{k,m}^{\textrm{align}}}(t) \right) \right) \wedge \omega_k(t)
	\end{array}\right. \;\forall k \in \llbracket 1,\textrm{N}_{\textrm{fib}} \rrbracket.
\label{eq_pfd}
\end{equation}

\subsection{Description of the experiments}
\label{section_modelexp}

\begin{table}
	\centering
	\begin{tabular}{|c|c|c|p{7cm}|}
		\hline Name & Value & Units & Description \\ \hline
		\multicolumn{4}{|c|}{\textbf{Agents}} \\\hline
		$\textrm{N}_{\textrm{fib}}$ & $[1500,3000]$ & N/A & Number of fibers \\\hline
		$L_{\textrm{fib}}$ & $6$           & $L$ & Fiber length \\\hline
		$R_{\textrm{fib}}$ & $0.5$         & $L$ & Fiber radius \\\hline
		\multicolumn{4}{|c|}{\textbf{Mechanical interactions}} \\\hline
	    $\alpha_{\textrm{rep}}$           & $12.5$ & $M.L^{-1}.T^{-2}$ & Magnitude of the repulsion force \\\hline
	    $\alpha_{\textrm{rest}}$          &  $5.0$ & $M.T^{-2}$        & Magnitude of the elastic restoring force \\\hline
	    $\alpha_{\textrm{align}}$         &  $2.0$ & $M.L^{2}.T^{-2}$  & Magnitude of the alignment torque \\\hline
		$d_{\textrm{link}}^{\max}$        &  $1.0$ & $L$               & Perception distance for link creation \\\hline
		$d_{\textrm{link}}^{\textrm{eq}}$ &  $1.0$ & $L$               & Link equilibrium length \\\hline
		\multicolumn{4}{|c|}{\textbf{Biological phenomena}} \\\hline
		$\nu_{\textrm{link}}$  & $[0,10]$    & $T^{-1}$ & Network remodeling speed \\\hline
		$\chi_{\textrm{link}}$ & $[0.1,0.9]$ & N/A      & Equilibrium linked fiber fraction    \\\hline
		\multicolumn{4}{|c|}{\textbf{Numerical parameters}} \\\hline
		$L_x = L_y = L_z$    & $30$     & $L$ & Side length of the cubic domain \\\hline
		$T_{\textrm{final}}$ & $5.10^4$ & $T$ & Total time of simulation        \\\hline
	\end{tabular}
	\caption{\centering Model parameters.}
    \label{table_parameters_simu}
\end{table}

We consider a spatial domain $\Omega$ which is a cuboid of side length $L_x$, $L_y$ and $L_z$ respectively in the $x$, $y$ and $z$-dimension, and is centered on the origin~:
$$ \Omega = \left[-\frac{L_x}{2},\frac{L_x}{2}\right] \times \left[-\frac{L_y}{2},\frac{L_y}{2}\right] \times \left[-\frac{L_z}{2},\frac{L_z}{2}\right]. $$

For the sake of simplicity, we assume periodic boundary conditions~: an agent exiting the domain by one side re-enters immediately from the opposite side, and interactions between agents are computed using the periodicized euclidean distance. In other words, $\Omega$ is topologically equivalent to the $3$D-torus.

We perform numerical simulations of our model on the domain $\left( \Omega \times \mathbb{S}_2^+ \right)^{\textrm{N}_{\textrm{fib}}}$ for various sets of parameters. Fibers are randomly inseminated inside the domain according to a uniform law for both position and orientation. The differential system~\eqref{eq_pfd} is then numerically solved using a discrete upwind Euler scheme with adaptive time-step, which has a very low computational cost. Further reduction of the computational cost is achieved by dividing $\Omega$ into cubes whose side-length is higher than the maximal range of the interactions~: thus, interactions need only be computed for pairs of agents located in neighbouring cubes. Details of the numerical implementation are given in annex~\ref{annex_numimpl}.

The physical scaling of all the parameters of the model, as well as the values used in the simulations, are described in Table~\ref{table_parameters_simu}. A few points may be noted~: the perception distance for link creation $d_{\textrm{link}}^{\max}$ and the link unloaded length $d_{\textrm{link}}^{\textrm{eq}}$ are both equal to their minimal acceptable value $2 R_{\textrm{fib}}$; the size of the domain is approximately $4$ times the size of a fiber along its main axis; the fiber aspect-ratio $\frac{L_{\textrm{fib}}}{2 R_{\textrm{fib}}} = 6$ is quite small compared to the values used in other models of the ECM, which usually varies between $250$ and $10^4$~\cite{Riz2012,Lee2014,Lin2015}. This compensate the fact that these models directly account for fiber bending and/or fiber elongation.

We denote by $\phi_{\textrm{fib}}$ the fiber density of the network, that is the ratio between the total volume of fibers (without overlapping) and the volume of the spatial domain~:
\begin{equation}
    \phi_{\textrm{fib}} = \frac{\textrm{N}_{\textrm{fib}} V_{\textrm{fib}}}{|\Omega|} = \textrm{N}_{\textrm{fib}} \frac{\pi R_{\textrm{fib}}^2 L_{\textrm{fib}} + (4/3) \pi R_{\textrm{fib}}^3}{L_x L_y L_z}.
\end{equation}

{The quantity} $\phi_{\textrm{fib}}$ can be compared to the packing density, that is the maximal fraction of the domain that can be occupied by densely packed fibers. When considering an ordered packing, the optimal configuration of sphero-cylinders is the same as that of cylinders and the resulting packing density is a weighted average of the packing densities of spheres and cylinders, which in our case gives $\phi_{\textrm{order}} = 0.89$. However, since in our model fibers are randomly inseminated, the situation is closer to what is called a random or amorphous packing~: particles are generated randomly with a volume exclusion constraint until it is no longer possible to inseminate another one. The density reached at that point is called the maximal random packing density $\phi_{\textrm{random}}$. For sphero-cylinders with an aspect-ratio of $6$, the literature gives us $\phi_{\textrm{random}} \approx 0.4$~\cite{Wil2003}.

Thus, we may say that a system is ``sparse'' if its fiber density is below $\phi_{\textrm{random}}$, ``dense'' if it is between $\phi_{\textrm{random}}$ and $\phi_{\textrm{order}}$, and ``hyperdense'' if it is above $\phi_{\textrm{order}}$. We study two types of systems~: dense systems containing $\textrm{N}_{\textrm{fib}} = 3000$ fibers ($\phi_{\textrm{fib}} = 0.58$) and sparse systems with $\textrm{N}_{\textrm{fib}} = 1500$ fibers ($\phi_{\textrm{fib}} = 0.29$).

For each of the three types of mechanical forces in the system, we define the ``characteristic interaction time'' the time needed for two isolated fibers interacting only via this force and initially positioned in the most unfavourable configuration to reach $99\%$ of the (asymptotic) equilibrium state. For repulsion, $T_{\textrm{rep}}$ is the time needed for two fully overlapped fibers ($\X_1 = \X_2$ and $\omega_1 = \omega_2$) to move apart by $99\%$ of their equilibrium distance $2 R_{\textrm{fib}}$ (i.e. $||\X_1 - \X_2|| = 0.99 \times 2 R_{\textrm{fib}}$). Similarly, for the elastic spring $T_{\textrm{rest}}$ is the time needed for two fibers that are initially fully overlapping and crosslinked at their center to move apart by $99\%$ of their equilibrium distance $d_{\textrm{link}}^{\textrm{eq}}$. On the other hand, for nematic alignment $T_{\textrm{align}}$ is the time needed for two perpendicularly intersecting fibers ($\X_1 = \X_2$ and $\omega_1 \perp \omega_2$) crosslinked at their center to reach a relative angle $\arccos(\omega_1 \cdot \omega_2) = 0.9^{\circ}$.

Explicit computation leads to the following formula (numerical values are given for the parameters presented in Table~\ref{table_parameters_simu})~:
\begin{equation}
    \left\lbrace\begin{array}{l} 
        \D T_{\textrm{rep}} = \frac{9 \mu_{\textrm{fib}} L_{\textrm{fib}}}{R_{\textrm{fib}}\ \alpha_{\textrm{rep}}} = 8.64 \, U_t, \\
        \D T_{\textrm{rest}} = \ln(100) \frac{\mu_{\textrm{fib}} L_{\textrm{fib}}}{\alpha_{\textrm{rest}}} = 5.53 \, U_t, \\
        \D T_{\textrm{align}} = 4.8 \frac{\mu_{\textrm{fib}} L_{\textrm{fib}}^3}{\alpha_{\textrm{align}}} = 523 \, U_t.
    \end{array}\right. 
\end{equation}

It may be noted that the alignment interaction is much slower than the repulsive and elastic restoring forces.

\section{Results}

\subsection{Matrix crosslinking drives the emergence of ordered structures in 3D dynamical networks.}
\label{section_corrNlinkAl}

\begin{figure}
    \centering
    \includegraphics[width=\textwidth]{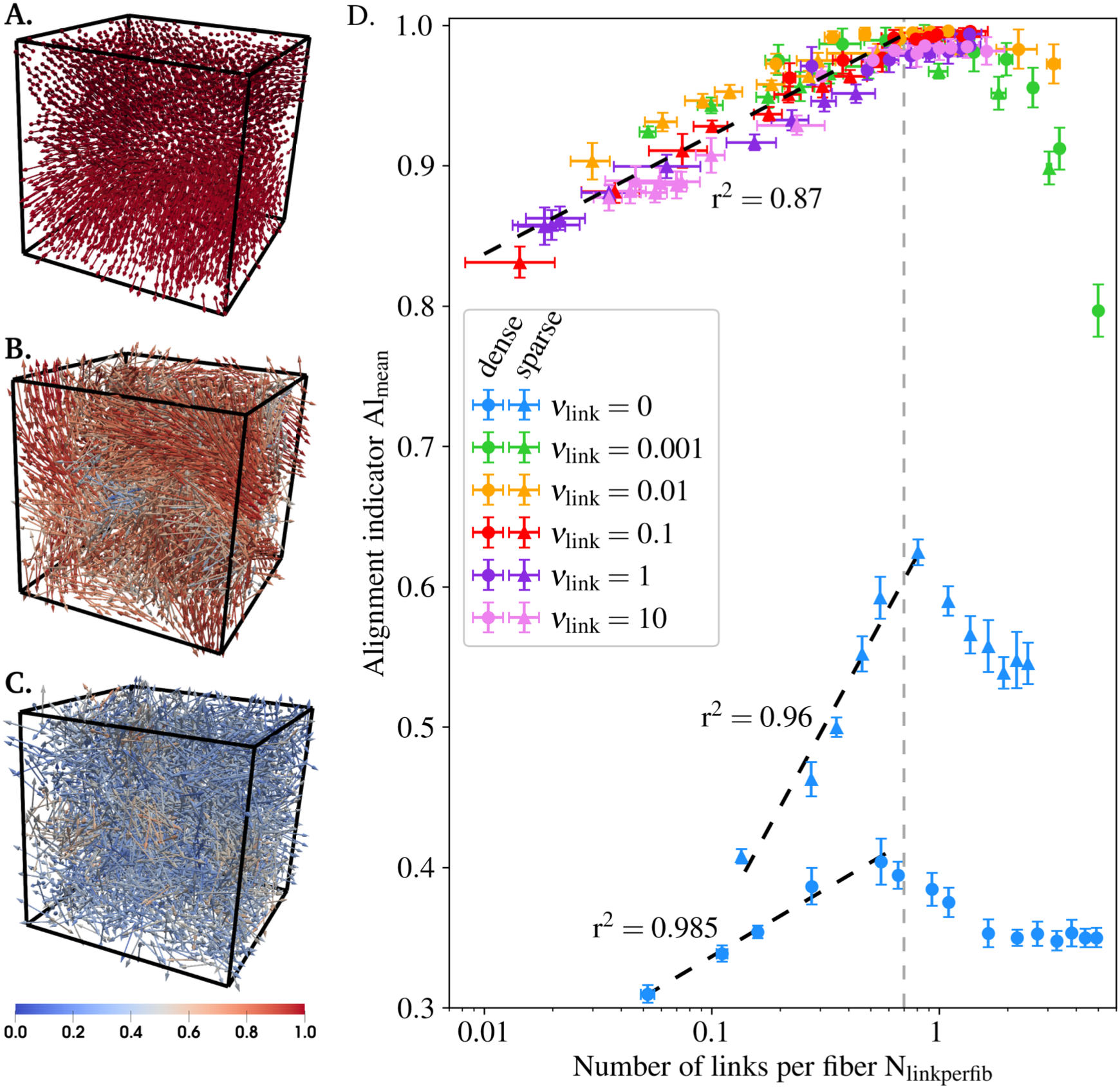}
    \caption{\textbf{Panels A-C:} Illustration of the various structures that can be observed at equilibrium. Fibers are represented by double-headed arrows and colored according to their local alignment with their neighbours (from blue~: $\textrm{Al}_{k} = 0$ to red~: $\textrm{Al}_{k} = 1$). The structures range from systems with uniformly high local alignment indicator (panel \textbf{A}) through systems with heterogeneous, intermediate local alignment indicator (panel \textbf{B}) to disordered systems with uniformly low local alignment indicator (panel \textbf{C}).
    \textbf{Panel D :} Value of $\textrm{Al}_{\textrm{mean}}$ according to $\textrm{N}_{\textrm{linkperfib}}$ at equilibrium, with color depending on the remodelling speed $\nu_{\textrm{link}}$. Each data-point represents the average value computed over $10$ simulations conducted with the same set of parameters, with horizontal and vertical error-bars for the standard deviation over $\textrm{N}_{\textrm{linkperfib}}$ and $\textrm{Al}_{\textrm{mean}}$ respectively. The gray dashed-line indicates the critical value of $\textrm{N}_{\textrm{linkperfib}}$ and the black dashed-lines the two logarithmic fits obtained for $\textrm{N}_{\textrm{linkperfib}} < \textrm{N}_{\textrm{critic}}$.}
    \label{fig_corrNlinkAl}
\end{figure}

In Figure~\ref{fig_corrNlinkAl}.(A-C), we show various structures that can be obtained with our model by playing on the parameters in the ranges indicated in Table~\ref{table_parameters_simu}. The fibers are represented by double arrows, colored as function of their local alignment with their neighbors. We refer the readers to annex~\ref{annex_quantif_al} for more details on the computation of this quantifier, and just mention that the local alignment of fiber $k$, denoted $\textrm{Al}_k$, is equal to $1$ (fiber colored in red) if all the neighbouring fibers display the exact same direction as fiber $k$, and to $0$ (fiber colored in blue) if the neighbouring fibers display uniformly distributed directional vectors. 

As one can observe, the fiber structures obtained at equilibrium range from highly aligned systems (mainly composed of red fibers, see Figure~\ref{fig_corrNlinkAl}.A) to disordered systems with a low local alignment (mainly composed of fibers colored in blue, see Figure~\ref{fig_corrNlinkAl}.C). The model can also produce intermediate states composed of fibers with a median local alignment (see Figure~\ref{fig_corrNlinkAl}.B). 

In order to assess the alignment states of our different fiber networks, we computed the mean and the standard deviation of the local alignment indicator $\textrm{Al}_k$ over all the fibers of the system, denoted by $\textrm{Al}_{\textrm{mean}}$ and $\textrm{Al}_{\textrm{STD}}$. By plotting this alignment quantifier $\textrm{Al}_{\textrm{mean}}$ (computed on the systems at equilibrium) as a function of the proportion of links $\textrm{N}_{\textrm{linkperfib}} = \textrm{N}_{\textrm{links}}/\textrm{N}_{\textrm{fib}}$, we discovered a striking and major correlation between these two quantities.

This correlation is shown in Figure~\ref{fig_corrNlinkAl}.D, where each point corresponds to the average over $10$ simulations conducted with the same set of parameters, with vertical and horizontal error-bars indicating $\textrm{Al}_{\textrm{STD}}$ and standard deviation of $\textrm{N}_{\textrm{linkperfib}}$ respectively. The different markers indicate different fiber densities (dots for dense systems and triangles for sparse ones), the different colors refer to different networks dynamics $\nu_{\textrm{link}}$, and inside each color series $\chi_{\textrm{link}}$ is increasing with $\textrm{N}_{\textrm{linkperfib}}$.

Figure~\ref{fig_corrNlinkAl}.D reveals that the values of $\textrm{Al}_{\textrm{mean}}$ and $\textrm{N}_{\textrm{linkperfib}}$ at equilibrium are highly correlated. When $\textrm{N}_{\textrm{linkperfib}}$ is inferior to a critical threshold $\textrm{N}_{\textrm{critic}} \approx 0.7$ (indicated with a grey dashed line on Figure~\ref{fig_corrNlinkAl}.D), there is a logarithmic correlation between the proportion of links in the network and its mean alignment indicator~:
\begin{equation}
    \textrm{Al}_{\textrm{mean}} \approx \alpha \log(\textrm{N}_{\textrm{linkperfib}}) + \beta,
\end{equation}
with 
\begin{itemize}
    \item $\alpha = 0.037$, $\beta = 1.006$ and coefficient of determination $r^2 = 0.87$ for dynamical systems (non-blue markers);
    \item $\alpha = 0.129$, $\beta = 0.651$ and coefficient of determination $r^2 = 0.96$ for sparse non-dynamical networks (blue triangles);
    \item $\alpha = 0.042$, $\beta = 0.433$ and coefficient of determination $r^2 = 0.985$ for dense non-dynamical networks (blue dots).
\end{itemize}
All these correlations are shown on Figure~\ref{fig_corrNlinkAl}.D with black dashed lines. 

Then, when $\textrm{N}_{\textrm{linkperfib}} > \textrm{N}_{\textrm{critic}}$ we observe an abrupt drop of the equilibrium value of $\textrm{Al}_{\textrm{mean}}$.

Surprisingly and very interestingly, for dynamical systems ($\nu_{\textrm{link}} > 0$) there is no difference in alignment induced by the fiber density or the link characteristics $\nu_{\textrm{link}}$ and $\chi_{\textrm{link}}$~: the correlation observed is the same for all sets of points.

The second major observation from Figure.~\ref{fig_corrNlinkAl}.D is the difference between non-dynamical and dynamical networks at equilibrium. Indeed non-dynamical networks, composed of a fixed number of links, are systematically less aligned than dynamical ones (compare the values of $\textrm{Al}_{\textrm{mean}}$ between the blue markers and the other colors). Moreover, although we do recover the same type of correlation between the fiber local alignment and the proportion of links in the network, for non-dynamical networks this correlation significantly depends on the fiber density. However, the critical number of links $\textrm{N}_{\textrm{critic}}$ allowing for larger alignment is the same for non-dynamical networks, either dense of sparse, and for dynamical networks. 

Altogether, these results show that the emergence of organized networks (i) requires some remodelling abilities of the ECM matrix and (ii) is mainly controlled by the proportion of its crosslinks.

Therefore, we performed a profound exploration of the role of the model parameters on the tissue architectures at equilibrium and in time, characterizing both the local arrangement of the fibers and the global architecture. In the following, we will take a particular focus on the role of matrix remodelling speed (viewed as a measure of its ``plasticity'').

\subsection{Characterization and quantitative assessment of various 3D architectures.}
\label{section_charac_states}

In this section, we take a step further in the qualitative and quantitative analysis of the various tissue architectures that emerge from our $3$D mathematical model. The goal is to describe as precisely as possible both the local organization of the fibers in the networks and the large-scale structures. 

First, by performing a theoretical analysis of the alignment quantifier $\textrm{Al}_k$ using different preset distributions of the fibers directional vectors (see Figure~\ref{fig_calibration_Al} in annex~\ref{annex_quantif_al}), we showed that it was able to discriminate between fibers located in randomly oriented environments (corresponding to $\textrm{Al}_k < 0.5$), fibers located in nearly planar environments (leading to $\textrm{Al}_k$ around $0.7$), and fibers located in nearly uni-directional environments (leading to $\textrm{Al}_k$ above $0.8$). 

Then, to observe the global distribution of the fibers, we used a stereographic projection of their directional vectors (see Figure~\ref{fig_stereoproj} in annex~\ref{annex_quantif_stereoproj} for a detailed explanation). Disregarding the spatial position of a fiber (that is, the position of its center), we represented its directional vector as a point on the surface of the unit half-sphere in $3$D and then projected it onto the unit disk in $2$D. The pole of the projection is chosen as the ``main direction'' of the system, that is the average of all directional vectors.

\begin{figure}
	\centering
    \includegraphics[width=\textwidth]{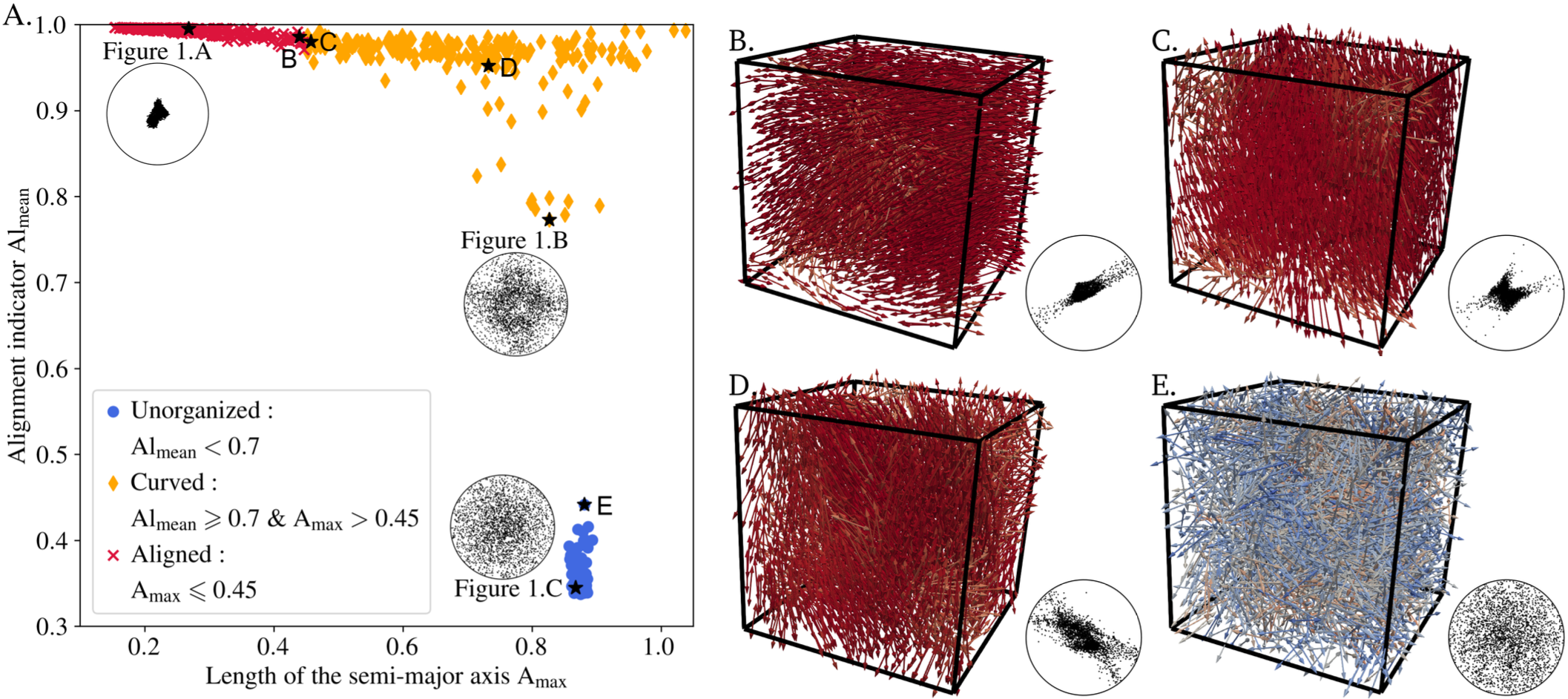}
    \caption{\textbf{Panel A}: Mean alignment indicator versus semi-major axis length of the covariance ellipse of the stereographic projection, for each simulation of dense systems ($\textrm{N}_{\textrm{fib}} = 3000$). Red crosses correspond to systems in an aligned state, orange diamonds to curved states and blue dots to unorganized states. The simulations previously displayed in Figure~\ref{fig_corrNlinkAl} are indicated with a black star and their stereographic projection given as inset. \textbf{Panels B to E} display the equilibrium state of a few other simulations, whose position on the diagram are also indicated with a black star.}
    \label{fig_phase_diagram}
\end{figure}

As shown in Figure.~\ref{fig_phase_diagram}, this representation enabled us to characterize the different global organizations of our fiber networks. Indeed, we observed three different types of stereographic projections in our simulations~: fibers directional vectors very concentrated around the center of the disk, corresponding to a global alignment of the system (Figure~\ref{fig_corrNlinkAl}.A and Figure~\ref{fig_phase_diagram}.A), fibers directional vectors homogeneously distributed on the disk corresponding to a global disorder (Figure~\ref{fig_corrNlinkAl}.C and Figure~\ref{fig_phase_diagram}.E), and fibers directional vectors distributed along a preferential axis, with complete depletion in the direction perpendicular to this axis, corresponding to global curved/plane structures (Figure~\ref{fig_phase_diagram}.(B-D)).

Therefore, the different states of our networks could be characterized both by a quantifier for local structuring such as $\textrm{Al}_{\textrm{mean}}$ and by quantifiers for global organization such as the size of the stereographic projection covariance ellipse $\textrm{A}_{\max}$. 

We considered a system to be locally aligned if the local distribution of its fibers directional vectors was mainly unidirectional, that is $\textrm{Al}_{\textrm{mean}}$ above $0.7$. At the same time, we considered that a system was globally aligned, in the sense that it displayed a global main direction, if its stereographic projection covariance ellipse had a semi-major axis smaller than $0.45$ (implying that the point cloud covers less than $20\%$ of the whole projection disk).

\begin{table}
    \centering
    \begin{tabular}{|c|m{1cm}|m{5cm}|m{5cm}|}\cline{3-4}
        \multicolumn{2}{c|}{} & \multicolumn{2}{c|}{$\textrm{A}_{\max}$} \\\cline{3-4}
        \multicolumn{2}{c|}{} & \multicolumn{1}{c|}{$\leqslant 0.45$} & \multicolumn{1}{c|}{$> 0.45$} \\\hline
        \multirow{4}{*}{$\textrm{Al}_{\textrm{mean}}$} & $\geqslant 0.7$ & \textbf{Aligned state :} alignment both local and global. & \textbf{Curved state :} alignment local but not global. \\\cline{2-4}
        & $< 0.7$ & (alignment global but not local) & \textbf{Unorganized state :} no alignment, either local or global. \\\hline
    \end{tabular}
    \caption{Classification of the simulations outcomes into different states based on the local quantifier $\textrm{Al}_{\textrm{mean}}$ and the global quantifier $\textrm{A}_{\max}$. The case $\{ \textrm{Al}_{\textrm{mean}} < 0.7 \;\&\; \textrm{A}_{\max} \leqslant 0.45 \}$ never occurs in our simulations and is thus unnamed.}
    \label{table_classification}
\end{table}

We therefore classified the simulations outcomes into three different states (unorganized, curved and aligned) using Table~\ref{table_classification}. We ran a total of $1080$ numerical simulations, exploring various values of the parameters $\nu_{\textrm{link}}$, $\chi_{\textrm{link}}$ and $\textrm{N}_{\textrm{fib}}$ in the broad ranges indicated in Table~\ref{table_parameters_simu}, and counted among their outcomes :
\begin{itemize}
    \item $180$ unorganized states (all occurring in non dynamical systems, i.e. $\nu_{\textrm{link}}=0$),
    \item $661$ curved states,
    \item $239$ aligned states (among which only $12$ occurred in sparse systems).
\end{itemize}

Figure~\ref{fig_phase_diagram}.A compare the values of quantifiers $\textrm{Al}_{\textrm{mean}}$ and $\textrm{A}_{\max}$ when the simulation has reached equilibrium, for dense systems (see annex~\ref{annex_snapshot_sparse} for the equivalent figure on sparse systems). The points are colored according to the states defined previously (blue dots correspond to unorganized states, orange diamonds to curved states and red crosses to aligned states). The simulations already displayed in Figure~\ref{fig_corrNlinkAl} are indicated with a black star and their stereographic projection shown as inset. Four other simulation outcomes are singled out with black stars and illustrated with a $3$D view and stereographic projection in the panels B to E on the right. Together, all these simulations give an overview of the various typical and borderline cases that can be generated by our model.

We first observe that the unorganized states (blue dots) form a small, compact group of points with large semi-major axis length while the aligned states (red crosses) make a long thin group with very high alignment indicator. On the other hand, the curved states (orange diamonds) form a scattered cloud of points with a broad range of values for both the semi-major axis length and the alignment indicator.

We can thus observe that the transition between unorganized and curved states is very sharp~: notice the gap between the blue dots and orange diamonds in panel A. Indeed, no simulation displays an average alignment indicator at equilibrium between $0.65$ and $0.77$ (including sparse systems, see annex~\ref{annex_snapshot_sparse}), and there is a marked difference between the least organized of the curved states (illustrated in Figure~\ref{fig_corrNlinkAl}.A) and the most organized of the unorganized states (illustrated in Figure~\ref{fig_phase_diagram}.E). This confirms our choice of $0.7$ for the threshold value between unorganized and curved states.

On the contrary, the transition from curved to aligned states is not a clear switch but a continuum of structures that can be illustrated by the two borderline cases in panels B and C of Figure~\ref{fig_phase_diagram}. Thus, one must be aware that the partition between curved and aligned states is partly arbitrary and depends on the choice of the threshold.

Our quantifiers therefore allowed to quantitatively characterize three different network architectures.

\subsection{ECM architecture emergence is driven by a complex interplay between remodelling speed and linked fiber fraction}
\label{section_impact_remodeling}

In this section, we use the quantifiers defined in previous section to study the impact of the model parameters on the different tissue architectures (aligned/curved/unorganized) obtained by the model. 

\begin{figure}
    \includegraphics[width=\textwidth]{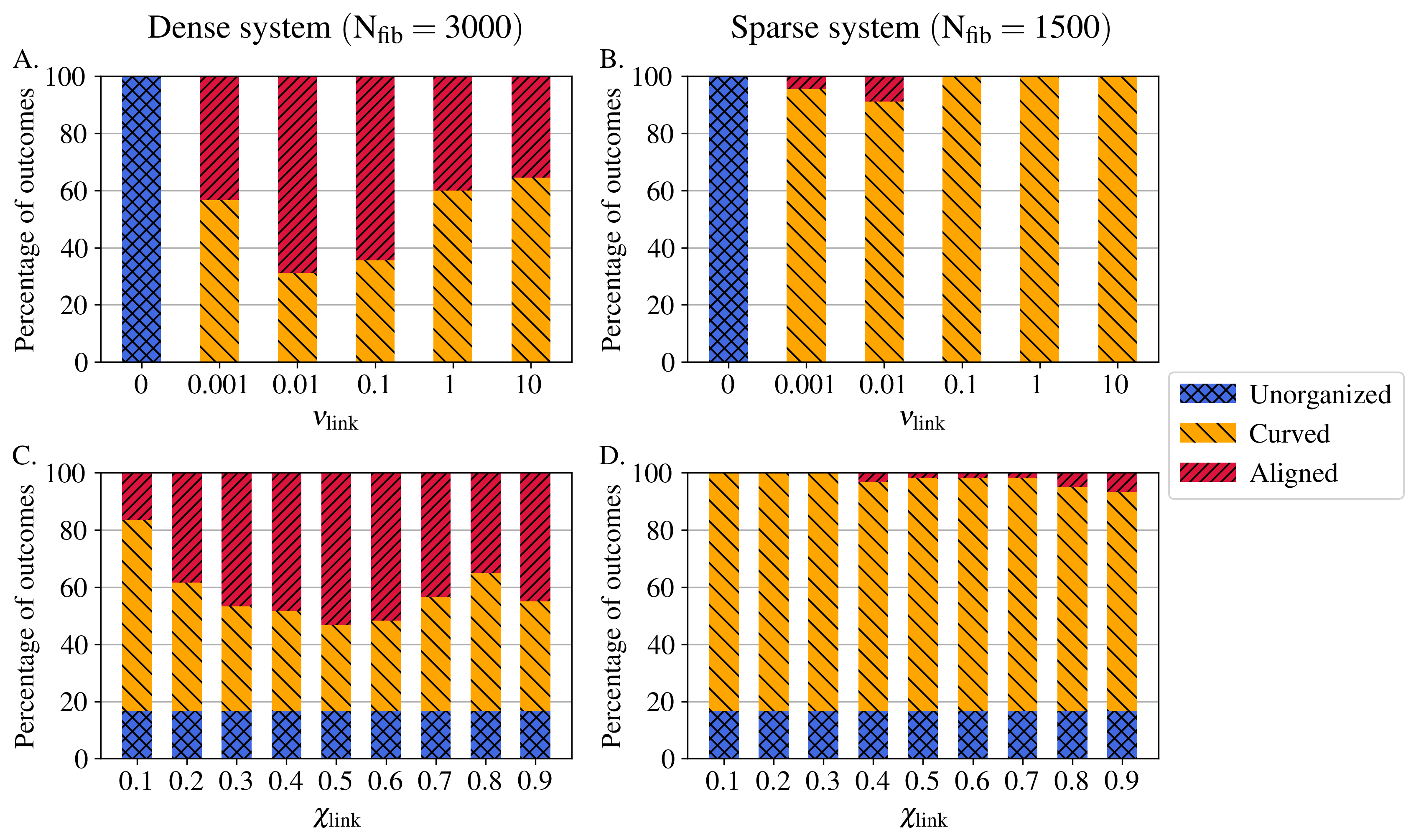} 
    \caption{Distribution of the outcomes of all $1080$ simulations between the different categories. Red zones correspond to systems in an aligned state, orange zones to curved states and blue zones to unorganized states. \textbf{Panels A and B}~: Each bar gives the percentage of each category among the outcomes of the $90$ simulations conducted with a given value of $\nu_{\textrm{link}}$ (on the x-axis) and $\textrm{N}_{\textrm{fib}}$ (dense for panel A and sparse for panel B). \textbf{Panels C and D}~: Each bar gives the percentage of each category among the outcomes of the $60$ simulations conducted with a given value of $\chi_{\textrm{link}}$ (on the x-axis) and $\textrm{N}_{\textrm{fib}}$ (dense for panel \textbf{C} and sparse for panel \textbf{D}).}
    \label{fig_corrParameters}
\end{figure}

In Figure~\ref{fig_corrParameters}, we show the distribution of the simulations outcomes at equilibrium, depending on the values of the network remodeling speed $\nu_{\textrm{link}}$ (panels A and B) and the equilibrium linked fiber fraction $\chi_{\textrm{link}}$ (panels C and D), for dense networks with $\textrm{N}_{\textrm{fib}}=3000$ (panels A and C) and sparse networks with $\textrm{N}_{\textrm{fib}}=1500$ (panels B and D). To account for the stochastic components of our model, we run $10$ simulations for each set of parameters. Thus, each bar in panels A and B represents which percentage of the $90$ simulations conducted with the indicated value of $\nu_{\textrm{link}}$ and $\textrm{N}_{\textrm{fib}}$ (varying over $9$ values of $\chi_{\textrm{link}}$) ended up in which state. And each bar in panels C and D represents which percentage of the $60$ simulations conducted with the indicated value of $\chi_{\textrm{link}}$ and $\textrm{N}_{\textrm{fib}}$ (varying over $6$ values of $\nu_{\textrm{link}}$) ended up in which state.

As already mentioned in section~\ref{section_corrNlinkAl}, we recover that non-dynamical networks ($\nu_{\textrm{link}} = 0$, left columns of Figures~\ref{fig_corrParameters}.A and B) are systematically unorganized, independently of the equilibrium linked fiber fraction or the fiber density. On the contrary, dynamical networks ($\nu_{\textrm{link}}>0$) never equilibrate in an unorganized state~: their plasticity (i.e. their ability to rearrange their connections) favours the formation of more organized states than non-dynamical networks. This shows that the discontinuous phase transition between unorganized and curved equilibrium states, revealed in Figure~\ref{fig_phase_diagram}, is controlled by $\nu_{\textrm{link}}$.

In contrast, the transition between the curved and aligned states is not controlled by a unique model parameter but is the interplay between several parameters. Indeed, we first observe in Figure~\ref{fig_corrParameters} that dense dynamical networks seem to have a greater ability to create aligned states than sparse networks, which tend to favour curved states (compare the red zones in panels A and C with the ones of panels B and D). Moreover, we also observe that, for both fiber densities, networks with a moderate remodeling speed $\nu_{\textrm{link}} \approx 0.01$ (middle column of panels A and B) seem to have a greater ability to reorganize into aligned states than low dynamical networks ($\nu_{\textrm{link}} \approx 0.001$) or fast remodeling networks ($\nu_{\textrm{link}} \ge 0.1$) (compare the red zones of each bar inside panels A and B). These results suggest that there exists a remodeling speed maximising the network alignment.

Looking at the impact of the equilibrium linked fiber fraction $\chi_{\textrm{link}}$, we observe different behaviours depending on the fiber density of the system. For sparse networks (Figure~\ref{fig_corrParameters}.D), increasing the equilibrium linked fiber fraction tends to favour a higher level of organization by increasing slightly the number of aligned states (red zones). On the contrary, dense networks (Figure~\ref{fig_corrParameters}.C) exhibit a more complex behaviour where intermediate fiber fraction $\chi_{\textrm{link}} \in [0.4,0.6]$ generate more aligned states (red zones), implying that there exists an equilibrium linked fiber fraction maximising the alignment of the system.

These results show that the different types of tissue architectures (aligned, curved or unorganized) depend on an interplay between parameters $\nu_{\textrm{link}}$ and $\chi_{\textrm{link}}$. While ECM local alignment can be explained by the simple emerging variable that is the proportion of links in the network (as shown in section~\ref{section_corrNlinkAl}), its direct relation with model parameters $\textrm{N}_{\textrm{fib}}$,  $\nu_{\textrm{link}}$ and $\chi_{\textrm{link}}$ is more complex. In the next section, we study the evolution in time of the structures, enabling to give more insights into the role of the parameters in tissue structuring.  

\subsection{ECM architecture emergence follows a unique evolutionnary path on timescales controlled by their remodelling characteristics}
\label{section_temporal_evol}

In this section, we study the temporal evolution of the spatial structures. 

Our very first observation is that, for all parameters, the quantifier $\textrm{Al}_{\textrm{mean}}$ follows an inverted exponential growth. We refer to annex~\ref{annex_temporal_evol_Almean} for a detailed analysis of $\textrm{Al}_{\textrm{mean}}$ as a function of time and do not show the curves here. We just mention here that we denote by $\tau_{\textrm{Al}}$ the time-constant of this growth, whose classical definition is the time needed for the quantifier to reach $63\%$ of its asymptotic value and which, in our case, correspond approximately to the time at which $\textrm{Al}_{\textrm{mean}}$ crosses the $0.7$ threshold between unorganized and curved states. This time-constant is a good time-scale to study the temporal evolution of the system and will be used as such in the following discussion.

\begin{figure}
    \centering
    \includegraphics[width=\textwidth]{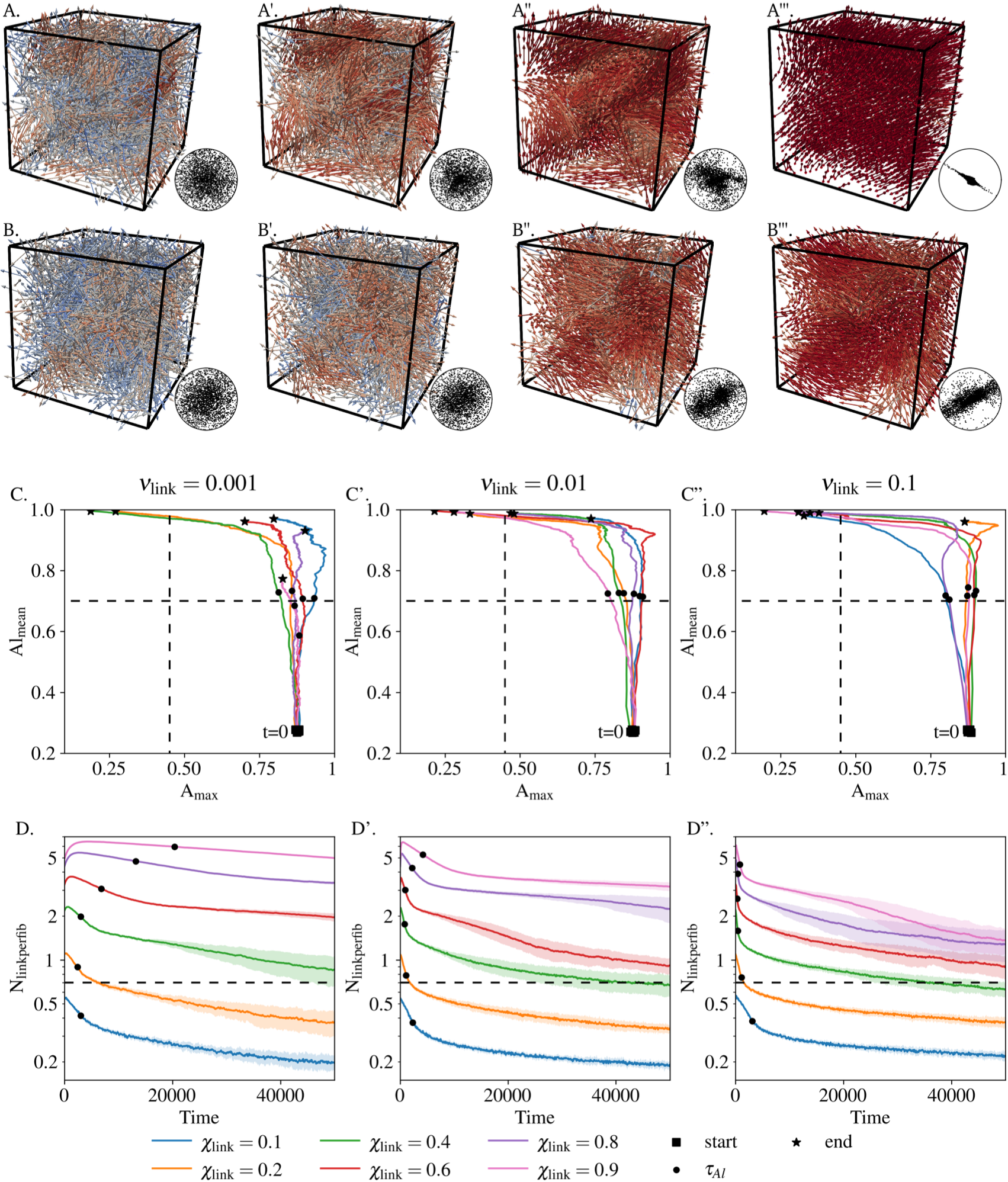}
    \caption{Temporal evolution of dense systems ($\textrm{N}_{\textrm{fib}} = 3000$) with various linking dynamics. \textbf{Panels A-A'''}~: $3$D view and stereographic projection of the system at times $0.5\tau_{\textrm{Al}}$ (\textbf{A}), $\tau_{\textrm{Al}}$ (\textbf{A'}), $3\tau_{\textrm{Al}}$ (\textbf{A''}) and $T_{\textrm{final}}$ (\textbf{A'''}), for one simulation with $\nu_{\textrm{link}}=0.1$ and $\chi_{\textrm{link}}=0.8$. \textbf{Panels B-B'''}~: $3$D view and stereographic projection of the system at times $0.5\tau_{\textrm{Al}}$ (\textbf{B}), $\tau_{\textrm{Al}}$ (\textbf{B'}), $3\tau_{\textrm{Al}}$ (\textbf{B''}) and $T_{\textrm{final}}$ (\textbf{B'''}), for one simulation with $\nu_{\textrm{link}}=0.001$ and $\chi_{\textrm{link}}=0.8$. \textbf{Panels C-C''}~: Trajectory in the phase plane $\textrm{A}_{\max}$ vs $\textrm{Al}_{\textrm{mean}}$ of a few individual simulations for different remodelling speeds $\nu_{link} = 0.001$ (\textbf{C}), $\nu_{link} = 0.01$ (\textbf{C'}) and $\nu_{link} = 0.1$ (\textbf{C''}). The initial position is indicated with a black square, the final position with a black star and the time-constant $\tau_{\textrm{Al}}$ with a black circle. The limits between each class of structures are drawn in dashed lines. \textbf{Panels D-D''}~: Evolution of $\textrm{N}_{\textrm{linkperfib}}$ for a few set of parameters. Each curve represents the average value computed over 10 simulations conducted with the same set of parameters, with shading indicating the standard deviation. The critical value $\textrm{N}_{\textrm{critic}}$ is indicated with a dashed line and the time-constant $\tau_{\textrm{Al}}$ with a black circle.} 
    \label{fig_time_trajectories_Nlink}
\end{figure}

Movies displaying the full temporal evolution of a few simulations are available in supplementary data (see annex~\ref{annex_movie}). In Figure~\ref{fig_time_trajectories_Nlink}.A-A''' and B-B''', we show the $3$D view and stereographic projection of a few well chosen time frames (namely $0.5\tau_{\textrm{Al}}$, $\tau_{\textrm{Al}}$, $3\tau_{\textrm{Al}}$ and $T_{\textrm{final}}$) for two of these simulations (respectively from \textit{Movie3.mp4} and \textit{Movie4.mp4}). They correspond to dense systems with $\chi_{\textrm{link}} = 0.8$ and two different crosslink dynamics~: fast remodeling network $\nu_{\textrm{link}} = 0.1$ (A-A''', \textit{Movie3.mp4}) and slow remodeling network $\nu_{\textrm{link}} = 0.001$ (B-B''', \textit{Movie4.mp4}). These screenshots enable us to answer the important question of how the network global structure emerges. It is not by accretion around a few structured areas that gradually merge together, but by an overall homogeneous structuring. Indeed, one can observe that the directional vectors gradually concentrate around a main direction without creating clustered points that merge together. This behavior can be observed both for very aligned networks (A-A''') or curved states (B-B'''), and in fact in all our simulations, independently on the network density. Therefore, our model suggests that the emergence of tissue architecture occurs on a global scale. \\

We now turn towards the analysis of the time trajectories of the spatial structures observed within our different networks. Figure~\ref{fig_time_trajectories_Nlink}.C-C'' displays the trajectory in the phase plane $\textrm{A}_{\max}$ vs $\textrm{Al}_{\textrm{mean}}$ of individual simulations conducted with various set of parameters. We chose this one-run representation instead of the usual $10$-runs average because the two quantifiers exhibit a non-negligible inter-simulations variability, so that plotting the standard deviation would blur the graphic but plotting only the average value would give a limited and partial view of the situation.

It can be seen that all the trajectories follow a common pattern. It begins with a sharp increase of the alignment indicator (from $0.15$ to between $0.4$ and $0.5$) while maintaining a quasi-constant semi-major axis length~: this corresponds to the partial depletion of one direction (denoted $d_1$) in the family of the fibers directional vector, thus shifting from the initial uniform distribution to a mainly two-directional distribution (see annex~\ref{annex_quantif_stereoproj} for more details on this interpretation). Non-dynamical networks (not represented on these graphics) do not go past that first stage.

The trajectories then diversify~: the alignment indicator keeps increasing while the semi-major axis length either decreases, stays constant or slightly increases. The first case is the most common and indicates that, while direction $d_1$ keeps depleting until near extinction, one of the two remaining directions starts to deplete as well. This diversification happens on the scale of the time-constant $\tau_{\textrm{Al}}$ of the alignment indicator (marked on the trajectories of Figure~\ref{fig_time_trajectories_Nlink}.C-C'' with a black circle).

Lastly, simulations ending in an aligned state and part of those ending in a curved state display a stage of condensation of the fibers directional vectors around a main direction. This is marked by a shrinking of the covariance ellipse and a slow increase of the alignment indicator, which has already nearly reached its steady state (compare with the stabilisation of $\textrm{Al}_{\textrm{mean}}$ in Figure~\ref{fig_time_Almean}). This last point comes from the local quality of the quantifier $\textrm{Al}_{\textrm{mean}}$~: a system can be very aligned locally, but not globally, if the main direction of the local structures varies smoothly across space. Thus, the transition between a curved and an aligned state is mostly characterized by a gradual shifting of multiple local structures towards the same direction, a phenomenon better registered by the quantifier $\textrm{A}_{\max}$ than $\textrm{Al}_{\textrm{mean}}$. \\

Finally, we observe that the number of links per fiber (displayed in Figure~\ref{fig_time_trajectories_Nlink}.D-D'') undergoes a transient increase followed by a two-stage exponential decay (appearing as a piece-wise linear decrease on the semi-logarithmic scale). The initial accumulation of crosslinks is more pronounced, in the sense that the peak is higher and the subsequent decrease slower, when $\chi_{\textrm{link}}$ is high and $\nu_{\textrm{link}}$ is low. For the extreme case of slow remodeling networks $\nu_{\textrm{link}} = 0.001$ with a large linked fiber fraction $\chi_{\textrm{link}} = 0.9$ (Figure~\ref{fig_time_trajectories_Nlink}.D), the phenomenon is so strong that only the first stage of exponential decay is observed during the time of the simulation. On the other hand, for fast remodeling networks ($\nu_{\textrm{link}} = 0.1$, Figure~\ref{fig_time_trajectories_Nlink}.D'') and/or small equilibrium linked fiber fraction ($\chi_{\textrm{link}} = 0.1$, blue curves), we do not observe any crosslinks accumulation.

This behaviour can be explained by comparing the linking dynamics to the characteristic time of the repulsive interaction $T_{\textrm{rep}} = 10$. Parameter $\chi_{\textrm{link}}$ describes the proportion of linked fibers among all linkable fibers at equilibrium, but this equilibrium takes time to establish (inversely proportional to $\nu_{\textrm{link}}$). If the repulsion interaction operates faster than the links remodeling (i.e. $T_{\textrm{rep}} \ll 1/\nu_{\textrm{link}}$), then the linkable configurations will change before the linking/unlinking processes could equilibrate on the current configuration~: new links will appear between newly overlapping fibers while former overlapping fibers will still be linked even if not overlapping anymore, leading to an accumulation of links in the system. This happens all the more if the disparity between the frequencies $\nu_{\textrm{link}}$ and $\nu_{\textrm{unlink}}$ is more favourable to linking than unlinking ($\nu_{\textrm{link}} > \nu_{\textrm{unlink}}$, i.e. if $\chi_{\textrm{link}} > 0.5$).

The system thus exhibits a global, macroscopic relaxation phenomenon which emerges from its various local, microscopic properties. It can be seen that the characteristic time-scale of this relaxation is comparable to the time-constant of the alignment indicator $\tau_{\textrm{Al}}$ (see position of the black circles on the curves in Figure~\ref{fig_time_trajectories_Nlink}.D-D'', which indicates the value of $\tau_{\textrm{Al}}$ for the corresponding set of parameters).

We conclude that slow remodeling networks with a high equilibrium linked fiber fraction $\chi_{\textrm{link}}$ first build up increasing stress and stiffen before slowly relaxing, while networks with low $\chi_{\textrm{link}}$ or fast remodeling networks exhibit stress relaxation and do not undergo high stiffening. As a result, these last types of networks reach higher local alignment at equilibrium.

These results demonstrate a nonlinear dependence of the network properties on the type and proportion of its crosslinks. A high number of long lasting crosslinks promotes crosslink accumulation resulting in medium/low alignment, while fast remodeling reduces the mechanical action of the individual links on the overall network, resulting in lowly connected networks being unable to align. The network alignment ability therefore requires a number of links adapted to its remodeling speed~: fast remodeling networks need a high equilibrium linked fiber fraction to quickly reach a high alignment indicator, while slow remodeling networks need a low equilibrium linked fiber fraction to prevent crosslink accumulation and the increase of matrix stiffness.

\section{Discussion}

In this work, we have implemented a $3$D model for fiber networks composed of fiber elements capable to dynamically crosslink or unlink each others, to align with each others at the crosslinks and to repel their nearest neighbors to prevent fibers from cluttering. We showed that this model can spontaneously generate various types of macrostructures whose emergence can be finely described. The model reveals that the different macrostructures (i) can be easily explained by a single emerging intermediate variable, namely the proportion of crosslinks in the ECM network, (ii) emerge homogeneously in space and not in a fragmented way, and (iii) follow the same unique evolutionary path for all structures and not multiple paths.

To our knowledge, this work is the first exhaustive study questioning the mechanisms of tissue architecture emergence via a simple mechanical model of dynamical fiber networks in $3$D. This framework reveals that the different tissue architectures at equilibrium is directly controlled by a simple intermediary variable, the proportion of links (see section~\ref{section_corrNlinkAl}). Our interpretation is that, when the number of links per fiber is inferior to the critical threshold $\textrm{N}_{\textrm{critic}}$, the network is weakly constrained. In this configuration, an increase in the number of links per fiber improves the transmission of information in the network and thus enhances the alignment process. The logarithmic scaling indicates that the higher the number of links per fiber, the less prominent this feature becomes, until the gain (in terms of the equilibrium alignment indicator) becomes null. The system then shifts into a constricted regime where each new link adds to the constriction of the network and impedes its reorganization, leading to a decrease of the local alignment. 

The fact that we observe the same correlation for all dynamical networks means that, as long as a network is slightly dynamical, its final alignment is mostly controlled by its proportion of links rather than by its remodelling dynamics or its density. On the other hand, non-dynamical networks are locked in mechanically constrained configurations, preventing the system from reorganizing efficiently compared to dynamical ones and leading to a much lower level of alignment. However, we showed that non-dynamical networks still contain some degrees of freedom allowing for spatial matrix reorganization, and that this organization is controlled again by the proportion of links in the network but also by the matrix density, which becomes an important factor. Indeed, denser networks are even less organized than sparse networks : this is due to the fact that denser networks are overcrowded, preventing any reorganization of their fibers.

The existence of a simple emerging variable such as the proportion of crosslinks to control tissue structuring can have major therapeutic implications in systems where the architecture of the ECM is impacted (scarring, fibrosis, ageing), but can also prove very useful in the field of tissue engineering. It is noteworthy that this variable is not prescribed by model parameters but emerges from the initial simple rules as a combination of ECM remodelling dynamics, linked fiber fraction and fiber spatial organization, independently of supplementary complex interactions involving external factors such as migrating cells, contractile forces etc. However, the correlation between crosslink proportion and fiber alignment only gives local information on the long-time structures (mean local alignment of the fibers at equilibrium).

The second major contribution lies in the analysis of the fine time evolution of the spatial structures. This documents the different temporal evolution of the structures as function of the ECM remodeling speeds and reveals an unique trajectory all architecture combined with internal and transient temporal windows during which they self-organize.

The equilibrium structures obtained with our model can be classified into three types~: (a) aligned states with a strong organization around one main direction, (b) curved states with a median, locally heterogeneous alignment indicator and a wide range of directional vectors living in a plane, named curved patterns and (c) unorganized states with very low alignment indicator and no preferential direction. Unorganized states were exclusively obtained for non-dynamical networks composed of permanent crosslinks ($\nu_{\textrm{link}} = 0$), whose plasticity was very low due to their inability to rearrange their crosslinks. In contrast, dynamical networks exhibited a mixture of aligned and curved states. These results point to the essential role of matrix remodeling in ECM structuring, consistent with several results in the literature (see~\cite{Bonnans2014} and references therein).

In emerging systems, the characteristics of the final outcome cannot be predicted from the initial rules of the system and the paths from the initial interactions to the final equilibrium can be numerous and complex corresponding to a stochastic evolution. This is not completely the case in our model because, if indeed the emerging macrostructures cannot be predicted from the initial rules and the emergence must be understood as a whole, the path is simple and unique and can be strongly predicted by an intermediate emerging variable (the proportion of crosslinks in the ECM). Our study suggests that the very aligned structures observed in fibrotic tissues could be mainly due to excess accumulation of crosslinks, consistent with the alterations of ECM structure observed as a consequence of increased crosslinking in lung fibrosis~\cite{Philp2018} or cancer~\cite{Levantal2009}, or again with previous studies on tissue-induced alignment of fibrous ECM~\cite{Piotrowski2017,Garrison2021}. Such deciphering of the emergence would open numerous perspectives for future investigations. Indeed, because of its simplicity, this emerging variable (the proportion of crosslinks in the ECM) is amenable for experimental measurements and represents a new putative target for the development of therapeutic drugs one could develop to restore the architecture of various biological tissues after external or internal alterations. In vivo experiments must be conducted to definitively validate this hypothesis and are out of the scope of this manuscript.

Finally, the temporal evolution of the structures revealed that dynamical networks composed of long-lasting links exhibited a phase of crosslink accumulation followed by a ``relaxation'' phase (reduction of the proportion of links in the network) associated with a spatial reorganization of its fibers, while fast remodeling networks exhibited only the ``relaxation'' phase. These results suggest possible mechanisms for crosslink accumulation observed for instance in ageing tissues~\cite{Wilson2013}. Moreover, the new insights into the temporal evolution of the structures as function of the ECM remodelling speed could prove useful in the field of tissue engineering, where there is a need to design efficient biological crosslinkers~\cite{Jiang2022,Xu2022}.

In this study, we demonstrated the ability of fiber networks to spontaneously self-organize as function of the kinetics of their crosslinks. It is noteworthy that our model features networks composed of only one type of crosslinks (permanent or transient with a given link-life). A natural perspective of our works would be to study the self-organization abilities of networks composed of heterogeneous crosslinks, following the works of~\cite{Mak2020}. Moreover, our network features active crosslinks, i.e crosslinks that generate an alignment of the fibers they are attached to. As a result, our fiber networks are not subject to any external mechanical stimuli. Future works will be devoted to the study of the mechanical properties of these dynamical networks under tensile/compressive stress, shear, etc. Another interesting perspective would be to add cells having the ability to generate locally biophysical cues such as tension, stiffness and fiber production/degradation and study these effects on the structure and mechanical properties of the ECM networks.

\appendix

\section{Model}
\subsection{Numerical implementation}
\label{annex_numimpl}

The differential system~\ref{eq_pfd} is numerically solved using a discrete upwind Euler scheme, with adaptive time-step. The linking and unlinking Poisson processes are updated between each time-step. We assume that a pair of fibers cannot change its linking state more than once in a single time-step~: this is reasonable if the length of the time-step $dt$ is small enough compared to the mean occurrence time $1/\nu$ of the Poisson process, so we prescribe $dt \le \textrm{dt}_{\textrm{link}}$ with
\begin{equation}
    \textrm{dt}_{\textrm{link}} = \min\left( \frac{0.5}{\nu_{\textrm{link}}} , \frac{0.5}{\nu_{\textrm{unlink}}} \right).
    \label{eq_dtlink}
\end{equation}

The probability for two fibers $k$ and $m$ to develop a crosslink between time $t_n$ and time $t_{n+1} = t_n + \textrm{dt}_n$ is then given by~:
\begin{equation}
    \mathbb{P}\left( p_{k,m}(t_{n+1}) = 1 \;\big|\; p_{k,m}(t_n) = 0 \text{ and } ||\X_{k,m}(t_n) - \X_{m,k}(t_n)|| \le d_{\textrm{link}}^{\max} \right) = 1 - e^{-\nu_{\textrm{link}} \textrm{dt}_n}
    \label{eq_proba_link}
\end{equation}
while the probability for a crosslink to break is given by~:
\begin{equation}
    \mathbb{P}\left( p_{k,m}(t_{n+1}) = 0 \;\big|\; p_{k,m}(t_n) = 1 \right) = 1 - e^{-\nu_{\textrm{unlink}} \textrm{dt}_n}
    \label{eq_proba_unlink}
\end{equation}

To ensure that agents do not swap position without even seeing each other, we also restrict the instantaneous translation of each fiber to half its radius $R_{\textrm{fib}}$ and its rotation to $\arctan(0.1) \approx 6^\circ$. This implies the following upper limits for the time-step~:
\begin{equation}
    \left\lbrace\begin{array}{l}
		\D \textrm{dt}_{\textrm{trans}}(t_n) = \min_{1 \le k \le \textrm{N}_{\textrm{fib}}}\left( 0.5 \frac{R_{\textrm{fib}}}{\left|\left| \frac{d\X_k}{dt}(t_n) \right|\right|} \right), \\[1em]
		\D \textrm{dt}_{\textrm{rot}}(t_n) = \min_{1 \le k \le \textrm{N}_{\textrm{fib}}}\left( \frac{0.1}{\left|\left| \frac{d\omega_k}{dt}(t_n) \right|\right|} \right).
    \end{array}\right.
    \label{eq_dtmove}
\end{equation}

Reduction of the computational cost is achieved by dividing the domain of simulation into cubes whose side-length is higher than the maximal range of the interactions~: thus, interactions need only be computed for pairs of agents located in neighbouring cubes. The loops calculating the interactions are parallelized for further speeding up of the simulations.

One iteration of the Euler scheme proceeds as follow~:
\begin{itemize}
    \item Parallel computation of all forces and torques sustained by the agents at time $t_n$ (right-hand part of equation~\eqref{eq_pfd}).
    \item Computation of the adaptive time-step (equations~\eqref{eq_dtlink} and~\eqref{eq_dtmove})
    $$\textrm{dt}_n = \min(\textrm{dt}_{\textrm{trans}}(t_n),\textrm{dt}_{\textrm{rot}}(t_n),\textrm{dt}_{\textrm{link}}).$$
    \item Motion of the agents to their new position~: 
    $$\X_k(t_{n+1}) = \X_k(t_n) + \textrm{dt}_n \frac{d\X_k}{dt}(t_n)$$ 
    $$\omega_k(t_{n+1}) = \omega_k(t_n) + \textrm{dt}_n \frac{d\omega_k}{dt}(t_n)$$
    \item Account for periodic boundary conditions.
    \item Attribution of each agent to a simulation box.
    \item Parallel update of linking configuration (equations~\eqref{eq_proba_link} and~\eqref{eq_proba_unlink}).
\end{itemize}

\subsection{Closest points of two finite segments}
\label{annex_closestpoints}

\begin{figure}
    \centering
    \includegraphics[width=0.5\textwidth]{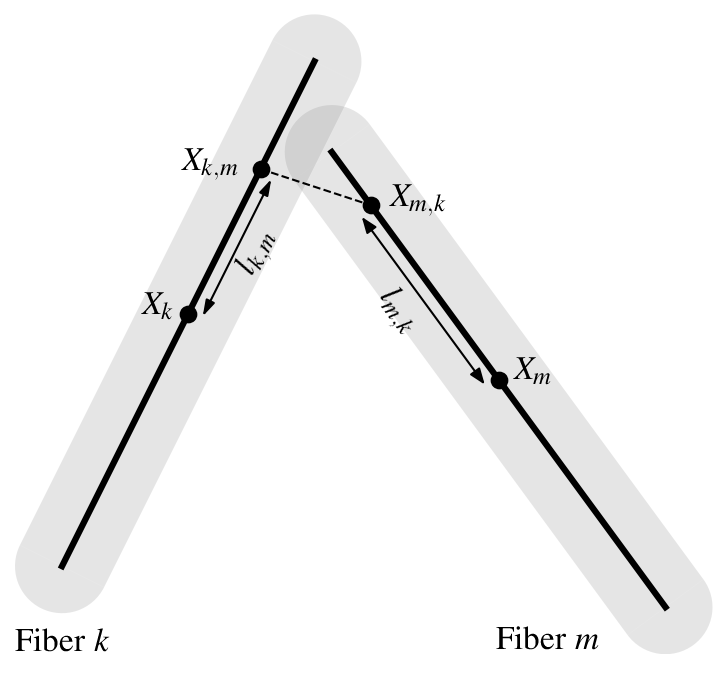}
    \caption{Scheme of two sphero-cylindrical fibers $k$ and $m$ indicating the position of the closest points $\X_{k,m}$ and $\X_{m,k}$ of their central segment (in a $3$D perspective) relative to their respective center.}
    \label{scheme_segment_closest}
\end{figure}

Given two fibers $k$ and $m$, we denote by $\X_{k,m} = \X_k + l_{k,m} \omega_k$ the point of fiber $k$ closest to fiber $m$ (see Figure~\ref{scheme_segment_closest}). The couple $(l_{k,m},l_{m,k})$ is the minimizer of the distance $||\X_k + u\omega_k - (\X_m + v\omega_m)||$ for $(u,v) \in \left[-\frac{L_{\textrm{fib}}}{2}, \frac{L_{\textrm{fib}}}{2}\right]$. If $\omega_k = \omega_m$, there is an infinity of solutions of the form $v = u + (\X_k - \X_m)\cdot\omega_k$; in this case we arbitrarily chose the solution with the smallest $|u|$ value. Otherwise, there exists a unique solution whose analytical expression is~:
\begin{equation}
    \left\lbrace\begin{array}{l}
        \D l_{k,m} = C_{ \frac{L_{\textrm{fib}}}{2} } \Big(\left( (\omega_k\cdot\omega_m) \omega_m \cdot (\X_k - \X_m) - \omega_k \cdot (\X_k - \X_m) \right) \big/ \left( 1 - (\omega_k\cdot\omega_m)^2 \right)\Big), \\[1em]
        \D l_{m,k} = C_{ \frac{L_{\textrm{fib}}}{2} } \Big(\left( (\omega_k\cdot\omega_m) \omega_k \cdot (\X_m - \X_k) - \omega_m \cdot (\X_m - \X_k) \right) \big/ \left( 1 - (\omega_k\cdot\omega_m)^2 \right)\Big),
    \end{array}\right.
\end{equation}
where $C_a$ denotes the cut-off function between $-a$ and $a$.

\section{Quantifiers and visualization tools for the fiber structures}
\label{annex_quantif}

\begin{figure}[ht]
    \centering
	\includegraphics[width=0.7\textwidth]{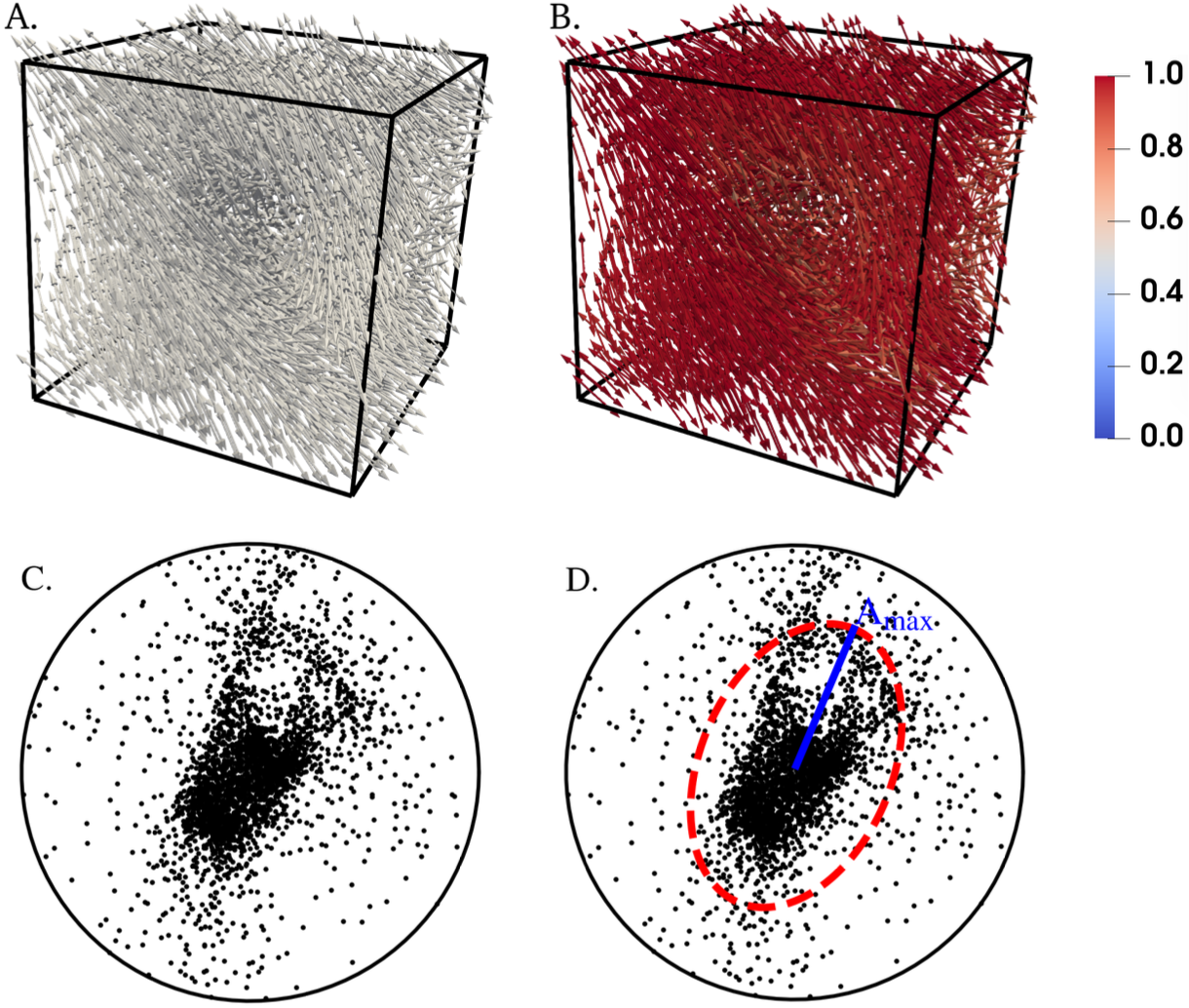}
    \caption{Illustration of the various way to visualize the state of a system, using as example the final state of a simulation. \textbf{Panel A}~: $3$D representation of each fiber as a gray double-headed arrow, with edges of the spatial domain $\Omega$ drawn in black. \textbf{Panel B}~: Same representation, with fibers colored according to their local alignment indicator (blue~: $\textrm{Al}_{k} = 0$, red~: $\textrm{Al}_{k} = 1$). See annex~\ref{annex_quantif} for the actual computation. \textbf{Panel C}~: Stereographic projection of the fibers directional vectors. See annex~\ref{annex_quantif_stereoproj} for the actual computation. \textbf{Panel D}~: Stereographic projection of the fibers directional vectors, with the covariance ellipse drawn in red dashed line and its semi-major axis drawn in blue solid line.}
    \label{fig_examplesimu}
\end{figure}

The goal of this section is to define quantifiers allowing to quantitatively describe the local and global organization of the fiber structures obtained with our computational model. Figure~\ref{fig_examplesimu}.A shows a typical simulation (almost) at equilibrium, in which fibers are represented as gray double arrows. As one can observe, this simulation shows two levels of organization~: a high local alignment and globally twisting, curving patterns located near the center of the domain. In order to quantitatively describe these states, we now define appropriate numerical quantifiers.

\subsection{Local alignment indicator}
\label{annex_quantif_al}

Let $R_{\textrm{align}}$ denotes the sensing distance up to which a fiber may interact with its neighbours~: in our model, it is equal to $L_{\textrm{fib}} + 2R_{\textrm{fib}}$. For any fiber $k$, we define its neighbourhood $\mathcal{B}_k$ as the set of all fibers located at a distance less than $R_{\textrm{align}}$ and its local alignment indicator $\textrm{Al}_k$ as the fractional anisotropy of the fibers directional vectors within $\mathcal{B}_k$.

It is computed as follows. We denote by $p_m = \omega_m \otimes \omega_m$ the projection matrix on the directional vector of fiber $m$. The mean of the projection matrices of the fibers inside $\mathcal{B}_k$ is given by
\begin{equation}
    P_k = \frac{1}{|\mathcal{B}_k|} \sum_{m \text{ s.t. } \X_m \in \mathcal{B}_k} p_m,
\end{equation}
where $|\mathcal{B}_k|$ denotes the number of fibers in $\mathcal{B}_k$.

The matrix $P_k$ is symmetric positive-definite, so its three eigenvalues $\lambda_1(P_k)$, $\lambda_2(P_k)$ and $\lambda_3(P_k)$ are real positive. The alignment indicator or fractional anisotropy in the neighbourhood $\mathcal{B}_k$ is then equal to~:
\begin{equation}
    \textrm{Al}_k = \sqrt{\frac{3}{2} \frac{ (\lambda_1(P_k) - \bar{\lambda})^2 + (\lambda_2(P_k) - \bar{\lambda})^2 + (\lambda_3(P_k) - \bar{\lambda})^2 }{ \lambda_1(P_k)^2 + \lambda_2(P_k)^2 + \lambda_3(P_k)^2 }}
\end{equation}
with $\bar{\lambda} = (\lambda_1(P_k) + \lambda_2(P_k) + \lambda_3(P_k))/3$ the mean of the eigenvalues.

Figure~\ref{fig_examplesimu}.B shows the same simulation as Figure~\ref{fig_examplesimu}.A, but here the fibers have been colored as a function of their local alignment indicator, from blue ($\textrm{Al}_k = 0$) to red ($\textrm{Al}_k = 1$). As one can see, the curved patterns are much easier to distinguish. Thus, the local alignment quantifier also serves as a visualization tool by supporting the qualitative, visual observation of locally organized states. \\

\begin{figure}
    \centering
    \includegraphics[width=0.8\textwidth]{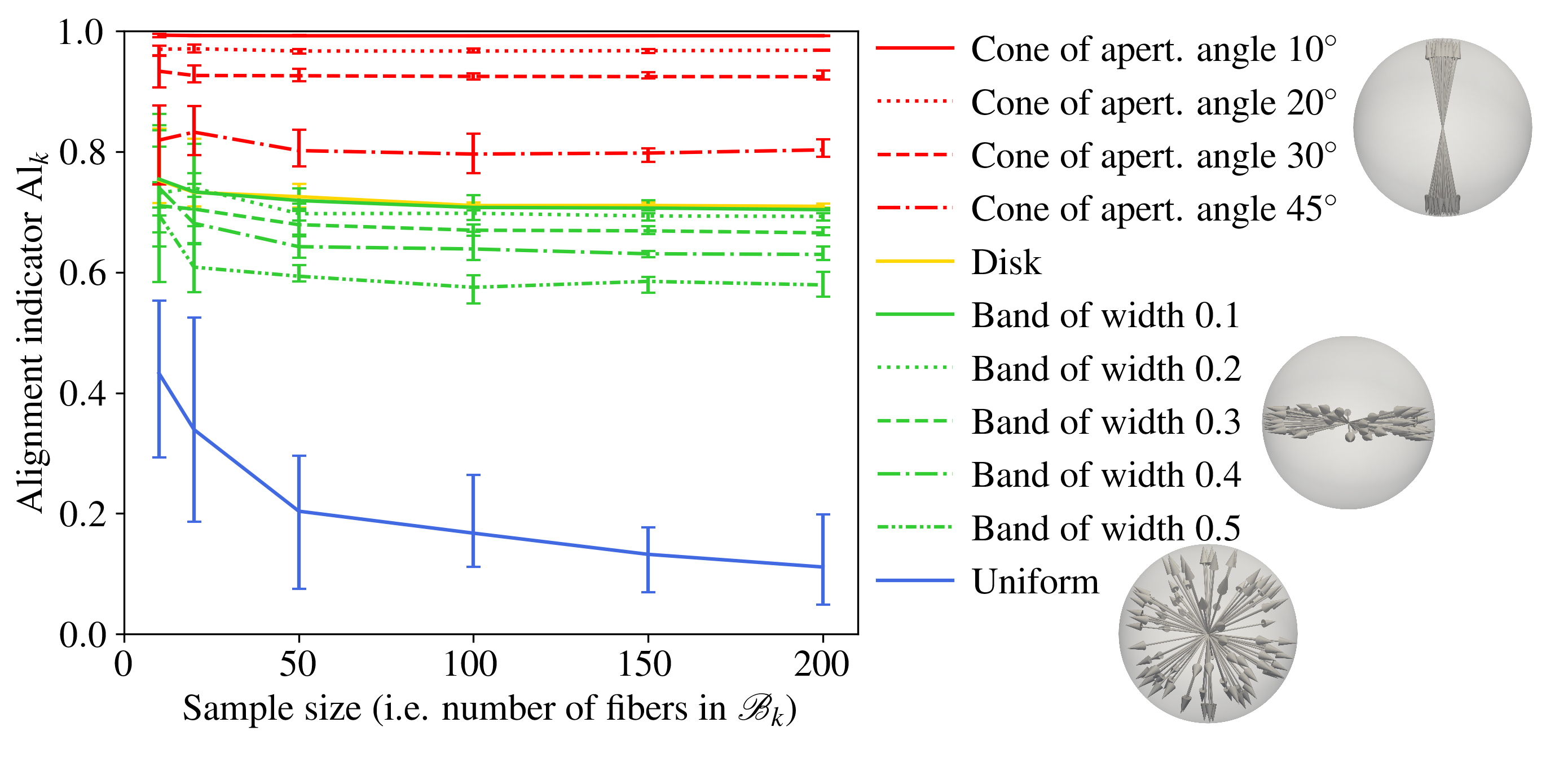}
    \caption{Calibration of the alignment indicator quantifier $\textrm{Al}$ on random sets of orientation vectors, for various distribution laws and sample sizes. The displayed values correspond to the average and standard deviation over $10$ random draws with the same characteristic.}
    \label{fig_calibration_Al}
\end{figure}

Note that $\textrm{Al}_k = 1$ if all the fibers in $\mathcal{B}_k$ have the same directional vector. If the directional vectors are uniformly distributed then theoretically $\textrm{Al}_k = 0$, but this is not always the case. Indeed, the actual sampling of a random distribution may not be fully isotropic, especially if the number of elements in the sample is small. Figure~\ref{fig_calibration_Al} displays the value of the alignment indicator obtained for various distribution of fibers and various sample sizes~: it can be seen that a uniform distribution produces alignment indicator ranging from $0.1$ (when the sample size is large) to as much as $0.55$ (when the sample size is small), and that there is a large discrepancy between different samples.

In our simulations, the number of neighbours of a fiber is very stable~: between $20$ and $25$ for dense systems and between $10$ and $15$ for sparse systems. Non-dynamical networks display mean alignment indicators between $0.3$ and $0.45$ for dense systems and between $0.4$ and $0.65$ for sparse systems~: these values are comparable to those observed in our calibration tests for a uniform distribution with similar sample size.

It can be seen from Figure~\ref{fig_calibration_Al} that these biases are much smaller for non-isotropic distributions~: for mainly two- or one-directional distributions, the values computed are nearly the same regardless of the sample size and the discrepancy between different samples is small. For a two-directional distribution (i.e. when the fiber directional vectors describe a disk), the eigenvalues on the mean projection matrix are theoretically $\lambda_1(P_k) = \lambda_2(P_k) = 1/2$ and $\lambda_3(P_k) = 0$, leading to a theoretical alignment indicator of $1/\sqrt{2} \approx 0.707$. This is very close to the value observed in our calibration tests (see yellow curve on Figure~\ref{fig_calibration_Al}). Nearly two-directional distributions, where the fiber directional vectors describe a ``band'' or thick disk, give lower and lower alignment indicator as the prominence of the third direction (i.e. the band width) increases (see green curves on Figure~\ref{fig_calibration_Al}). Likewise, conical distributions, which are mainly one-directional, give an alignment indicator close to $1$ which becomes lower and lower as the aperture angle of the cone increase (see red curves on Figure~\ref{fig_calibration_Al}).

\subsection{Stereographic projection}
\label{annex_quantif_stereoproj}

\begin{figure}
    \centering
    \includegraphics[width=\textwidth]{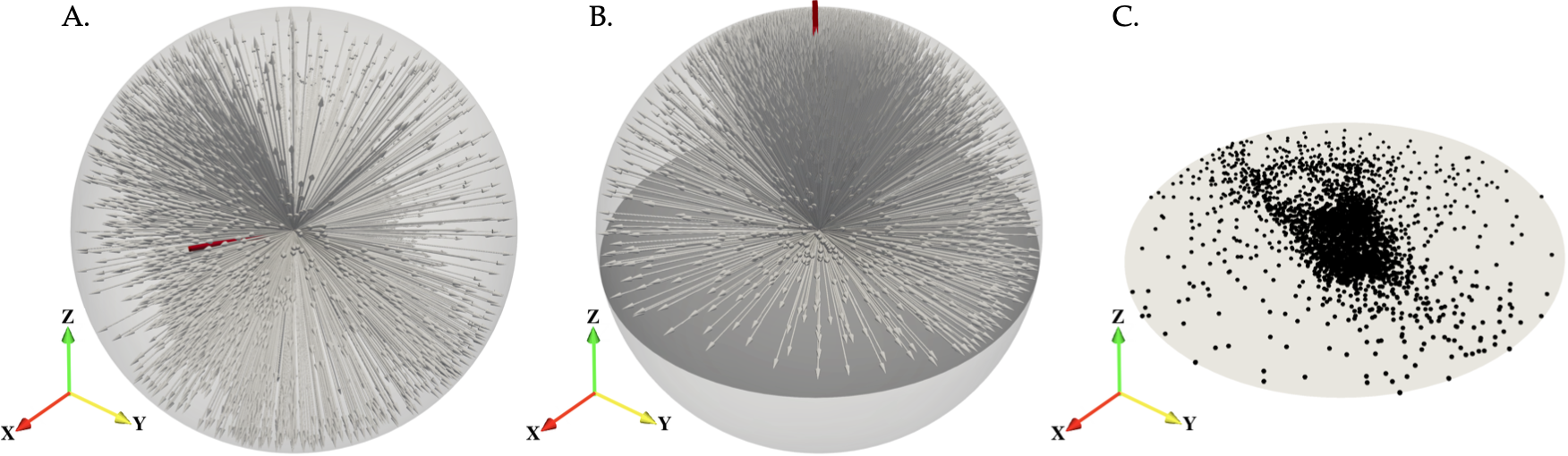}
    \caption{Illustration of the stereographic projection. The orientation axis are shown for reference. \textbf{Panel A}~: Natural distribution of the fibers directional vectors on the unit sphere $\mathbb{S}_2$, with main direction indicated by a red line. \textbf{Panel B}~: Rotation of the vectors set so that its main direction (in red) now lies along the $z$-axis. The definition-space of the vectors have been reduced to the ``north hemisphere'', that is to the subset $\mathbb{S}_2^+$ in the new rotated coordinates system. The equatorial plane is shown in dark-grey. \textbf{Panel C}~: Projection of the vectors onto the equatorial plane, shown in $3$D perspective.}
    \label{fig_stereoproj}
\end{figure}

The directional vectors of the fibers belong to the half unit sphere $\mathbb{S}_2^+$. This subset of $\mathbb{R}^3$ can be projected onto the unit disk in $2$D using a stereographic projection, as explained below.

We define the main direction of a system as the eigenvector associated to the largest eigenvalue of its total projection matrix
\begin{equation}
    P_{\textrm{tot}} = \frac{1}{\textrm{N}_{\textrm{fib}}} \sum_{1 \le k \le \textrm{N}_{\textrm{fib}}} \omega_k \otimes \omega_k.
\end{equation}

If the system contains two or three equally represented directions (associated to equal eigenvalues), one of them is randomly selected.

We rotate the set of directional vectors so that this main direction lies on the $z$-axis or ``north-south axis''. Since the fibers orientation is not relevant in our model, the set of directional vectors can be restricted to the ``north hemisphere'' of the sphere. A point $\omega = (x,y,z)$ on this hemisphere can then be projected onto the equatorial plane via the following transformation~:
\begin{equation}
    p(\omega) = \left( \frac{x}{1+z} , \frac{y}{1+z} \right).
\end{equation}

The whole process is illustrated in Figure~\ref{fig_stereoproj}. \\

Figure~\ref{fig_examplesimu}.C shows the stereographic projection of the simulation displayed in Figure~\ref{fig_examplesimu}.A and B. As one can observe, the dots are not uniformly distributed but densely packed at the center of the figure, indicating the existence of a main preferential direction in the system. However, not all fibers have a directional vector close to this main direction~: a non negligible number of dots are distributed all around the circle, meaning that all possible directions are represented in the system. Furthermore, the presence of a ``circular branch'' in the top-right part of the point cloud allows to identify the locally twisting structure that can be observed in Figure~\ref{fig_examplesimu}.B~: in this part of the system, nearby fibers have similar but gradually shifting directional vectors such that, on the scale of the whole structure, the fibers directional vectors describe a circle (in the domain $\mathbb{S}_2^+$).

Thus, this representation enables us to quickly grasp the distribution of the fibers directional vectors around one or more poles. It must be noted that proximity on the stereographic projection indicates similar directional vectors, but not necessarily spatial proximity. Nonetheless, we can gain insights into the overall architecture of the network by drawing the covariance ellipse of the point-cloud (in red dashed line on Figure~\ref{fig_examplesimu}.C) and computing its semi-major axis length $\textrm{A}_{\max}$. As shown in the section~\ref{section_charac_states}, this enables us to identify many type of ``states'' or structures that can also hint on the spatial organization of the network.

\section{Supplementary data}

\subsection{Temporal evolution of the spatial structures~: Movies}
\label{annex_movie}

We describe here the movies showing the results of some of our simulations, available online at \href{https://github.com/chassonnery/3D\_DynamicalFiberNetwork/tree/e99a4d07669c6b44fd55f5a866e8826da41d7a15/Supplementary%20data%20-%20Movies}{this address}. 

Each video is divided in three panels. On the left is a $3$D representation of the system with fibers colored according to their alignment indicator (see colorbar on the right) and edges of the spatial domain drawn in black, in the middle the stereographic projection of the fibers directional vectors and on the right the trajectory of the simulation in the plane $\textrm{A}_{\max}$--$\textrm{Al}_{\textrm{mean}}$. The current time (in $U_t$) is displayed at the top. \\

\textit{(Movie1)} Simulation of a dense system ($L_{\textrm{fib}} = 3000$) with fast remodeling dynamics ($\nu_{\textrm{link}} = 0.1$) and low equilibrium linked fiber fraction ($\chi_{\textrm{link}} = 0.2$).

This video shows a system quickly organizing~: at $t = 1000 U_t$, the system has already transitioned from its initial unorganized state to a curved state. At $t = 3000 U_t$, the main direction can be seen emerging in the form of a large cluster of points in the stereographic projection. At $t= 10\ 000 U_t$, the stereographic projection displays a planar distribution of the directional vectors, with extra accretion of points in the main direction and total depletion in the perpendicular direction. The system has already nearly reached its maximal alignment indicator and, from that point onward, it will mainly undergo small local adjustments of the fibers position and orientation (see the $3$D representation on the left panel). The alignment indicators of individual fibers harmonise, the mean alignment indicator increases slightly and the point cloud of the stereographic projection condensates in a clear straight band. During the entire simulation, the semi-major axis length $\textrm{A}_{\max}$ of the stereographic projection covariance ellipse stays nearly constant. \\

\textit{(Movie2)} Simulation of a dense system ($L_{\textrm{fib}} = 3000$) with slow remodeling dynamics ($\nu_{\textrm{link}} = 0.001$) and low equilibrium linked fiber fraction ($\chi_{\textrm{link}} = 0.2$).

This video shows a system organizing more slowly than the previous one (approximately twice slower) but achieving a more aligned final state. The system reaches a curved state at $t = 1900 U_t$. The main direction can be seen emerging on the stereographic projection around time $t \approx 5000 U_t$. The point cloud of the stereographic projection then begins to condensate around this main direction in a nearly symmetric manner while the various local structures rotate to align together (see left panel), reaching an aligned state at $t = 23\ 000 U_t$ and continuing to align. \\

\textit{(Movie3)} Simulation of a dense system ($L_{\textrm{fib}} = 3000$) with fast remodeling dynamics ($\nu_{\textrm{link}} = 0.1$) and high equilibrium linked fiber fraction ($\chi_{\textrm{link}} = 0.8$).

This video shows a system organizing very quickly, with a stereographic projection adopting as early as $t = 4000 U_t$ a band-like pattern which quickly gets thinner. At $t = 6000 U_t$, the $3$D representation shows a clear wavy pattern with very uniform local alignment indicators. At that time the mean alignment indicator is already high ($> 0.9$). The stereographic projection then begins to contract while the wavy pattern flatten, and the simulation ends in an aligned state. \\

\textit{(Movie4)} Simulation of a dense system ($L_{\textrm{fib}} = 3000$) with slow remodeling dynamics ($\nu_{\textrm{link}} = 0.001$) and high equilibrium linked fiber fraction ($\chi_{\textrm{link}} = 0.8$).

This video shows a system evolving very slowly. The mean alignment indicator reaches the $0.5$ threshold around $t = 11\ 000 U_t$. At that time, the local alignment indicator of individual fibers displays wide discrepancies and the stereographic projection point cloud has not visibly changed. A main direction can be seen emerging at approximately $t = 20\ 000 U_t$, but the central cluster of points is very large and does not contract over time, as can be seen by the fact that the quantifier $\textrm{A}_{\max}$ does nearly not decrease. The system ends in a curved state with heterogeneous local structures. \\

\textit{(Movie5)} Comparison of two simulations with different fiber densities, both ending in an aligned state. The top row shows a dense system ($L_{\textrm{fib}} = 3000$) with intermediate remodeling dynamic ($\nu_{\textrm{link}} = 0.01$) and moderate equilibrium linked fiber fraction ($\chi_{\textrm{link}} = 0.3$). The bottom row displays a sparse system ($L_{\textrm{fib}} = 1500$) with intermediate remodeling dynamic ($\nu_{\textrm{link}} = 0.01$) and very high equilibrium linked fiber fraction ($\chi_{\textrm{link}} = 0.9$).

It is noteworthy that the two systems display a very similar temporal evolution. This comes from the fact that they have the same remodeling speed $\nu_{\textrm{link}}$ and a comparable number of links per fiber $\textrm{N}_{\textrm{linkperfib}}$. The latter is achieved by giving the sparse system a higher equilibrium linked fiber fraction $\chi_{\textrm{link}}$, which compensate for its lesser number of linkable configurations (i.e. overlapping fiber pairs).

\subsection{Snapshots of sparse systems}
\label{annex_snapshot_sparse}

\begin{figure}
	\centering
    \includegraphics[width=\textwidth]{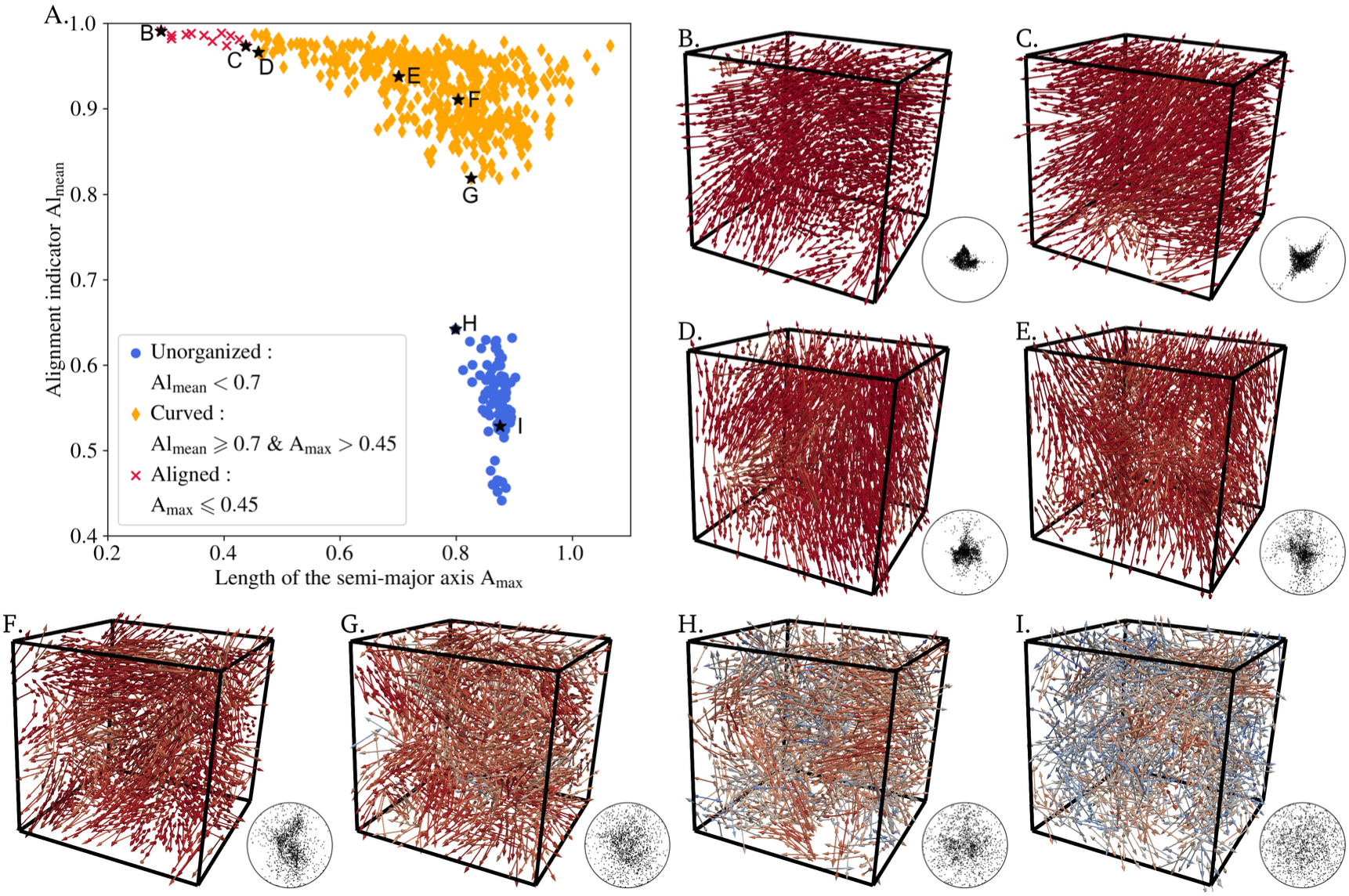}
    \caption{\textbf{Panel A:} Mean alignment indicator versus semi-major axis length of the covariance ellipse of the stereographic projection, for each simulation of a sparse system. Red crosses correspond to systems in an aligned state, orange diamonds to curved states and blue dots to unorganized states. \textbf{Panels B to I} display the equilibrium state of a few simulations (with $3$D view and stereographic projection) to illustrate typical or borderline cases. Their position on the diagram are indicated with a black star.}
    \label{fig_phase_diagram_sparse}
\end{figure}

In this section, we take a closer look at the spatial organization of sparse systems. Figure~\ref{fig_phase_diagram_sparse}.A compares the values of quantifiers $\textrm{Al}_{\textrm{mean}}$ and $\textrm{A}_{\max}$ when the simulation has reached equilibrium, with color depending on the type of state reached (blue dots correspond to unorganized states, orange diamonds to curved states and red crosses to aligned states). A few of simulations corresponding to either typical or borderline cases are singled out with black stars and illustrated with a $3$D view and stereographic projection in the panels B to I.

We first observe that the group of unorganized states (blue dots) is less compact than it was for dense systems and reaches greater values of $\textrm{Al}_{\textrm{mean}}$. The groups of curved states (orange diamonds) and aligned states (red crosses) have the same characteristics in term of $\textrm{Al}_{\textrm{mean}}$ and $\textrm{A}_{\max}$ than before, but the first one is much more populated and the second much less (it only contains $10$ simulations).

The most aligned state observed in sparse systems (panel B) is less straight than the typical aligned state for dense systems. Typical curved states (panels E and F) and unorganized states (panel I) however are very comparable to what was observed in dense systems. The transition between aligned and curved states is still continuous (compare panels C and D) and the transition between curved and unorganized states sharp (compare panels G and H), though the gap (in term of $\textrm{Al}_{\textrm{mean}}$) and the visual difference are lesser.

\subsection{Correlation between the links life-expectancy and the ECM architecture}
\label{annex_impact_Tlink}

\begin{figure}
    \centering
    \includegraphics[width=\textwidth]{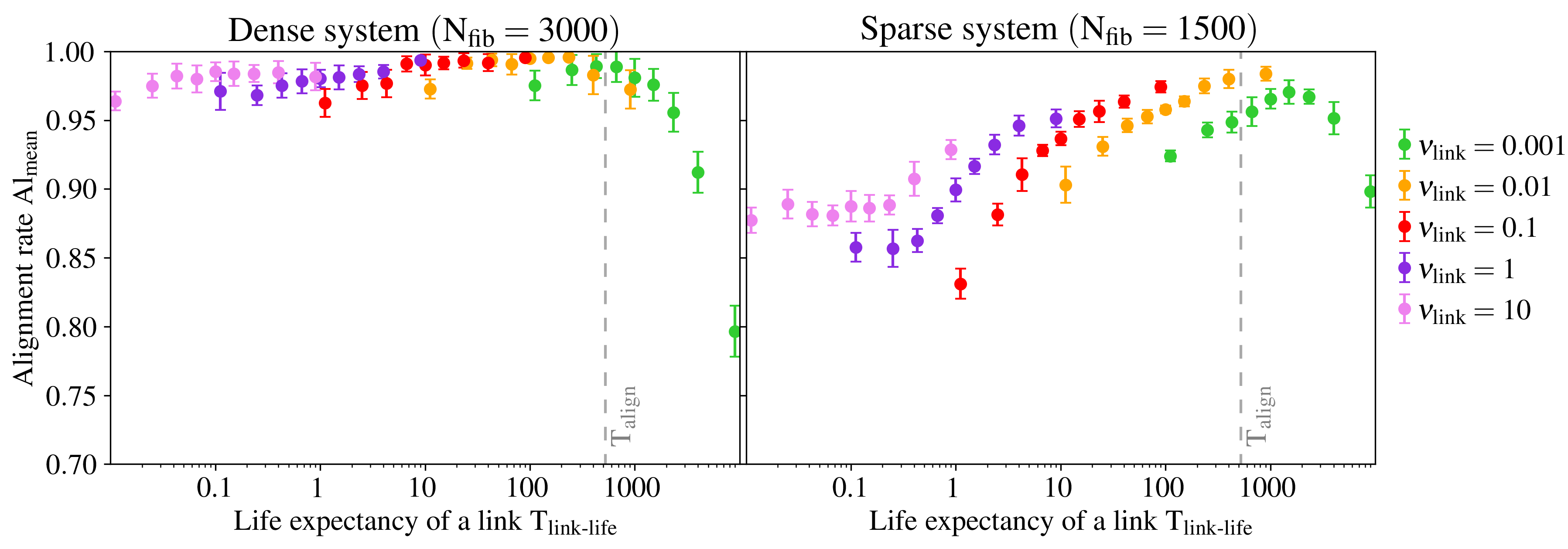}
    \caption{Value of $\textrm{Al}_{\textrm{mean}}$ at final time according to the value of $T_{\textrm{link-life}}$, with color depending on the remodeling speed $\nu_{\textrm{link}}$. The displayed values correspond to the average and standard deviation over $10$ simulations conducted with the same set of parameters. The characteristic time of the alignment interaction $T_{\textrm{align}}$ is indicated with gray dashed-lines for the sake of comparison.}
    \label{fig_corrTlink}
\end{figure}

Here, we explore whether the network organization abilities could be controlled by the life expectancy of a link, which depends of both $\nu_{\textrm{link}}$ and $\chi_{\textrm{link}}$ via the following relation~:
\begin{equation}
    T_{\textrm{link-life}} = \frac{1}{\nu_{\textrm{unlink}}} = \frac{\chi_{\textrm{link}}}{(1 - \chi_{\textrm{link}}) \nu_{\textrm{link}}}.
    \label{eq_Tlink}
\end{equation}

Figure~\ref{fig_corrTlink} displays the value of $\textrm{Al}_{\textrm{mean}}$ at equilibrium as a function of $T_{\textrm{link-life}}$. Each point corresponds to the average over $10$ simulations conducted with the same set of parameters, with a vertical error-bar indicating the standard deviation. The value of $\nu_{\textrm{link}}$ is indicated in color and, inside each color series, $\chi_{\textrm{link}}$ is increasing with $T_{\textrm{link-life}}$. The characteristic time of the alignment interaction $T_{\textrm{align}}$ (see section~\ref{section_modelexp}) is indicated for the sake of comparison.

We observe that, in the case of dense systems (left panel), the alignment indicator displays a flat maximum for $T_{\textrm{link-life}} \in [10, 500]\ U_t$, while for sparse system (right panel) it reaches its highest value at $T_{\textrm{link-life}} \approx 500\ U_t$. 

This can be explained by the fact that, when the average life expectancy of a link $T_{\textrm{link-life}}$ is very small compared to the characteristic time of the alignment force $T_{\textrm{align}} = 523\ U_t$, the links do not persist long enough to fully exert their aligning influence and the equilibrium alignment indicator is lesser. This is especially true for sparse systems, which display a clear drop for $T_{\textrm{link-life}} < 500\ U_t$. For dense systems the drop is slower and less pronounced.

On the other hand, when $T_{\textrm{link-life}}$ is large compared to $T_{\textrm{align}}$, on average the links last longer than necessary to wield their full effect and lock the system in non-optimal configurations by obstructing the action of other links. Though these locally locked structures will disappear over time, others will appear - or, to put it another way, the transmission of information (i.e. the fiber direction) in the network is too slow for all the agents to synchronize and the system will not be able to reach an extremely aligned equilibrium state.

\subsection{Temporal evolution of quantifier $\textrm{Al}_{\textrm{mean}}$}
\label{annex_temporal_evol_Almean}

\begin{figure}
    \centering
    \includegraphics[width=\textwidth]{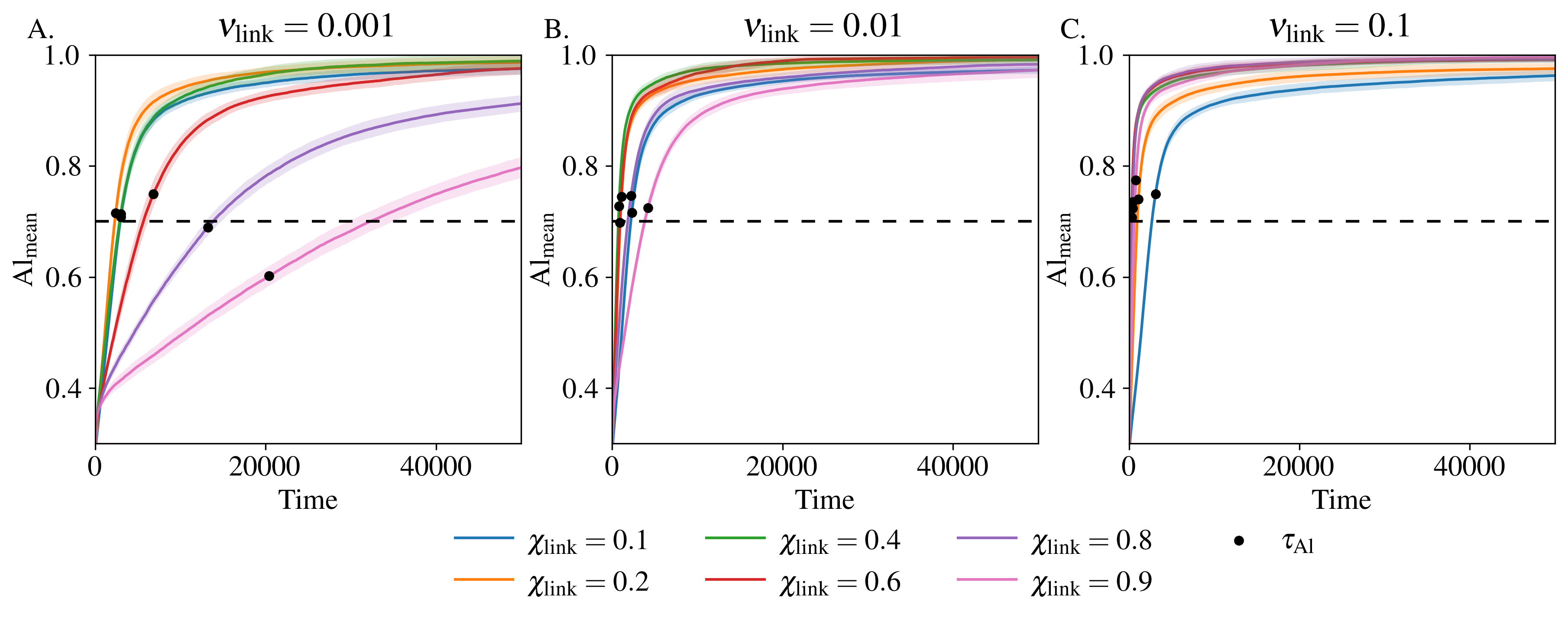}
    \caption{Temporal evolution the quantifier $\textrm{Al}_{\textrm{mean}}$ for dense systems ($\textrm{N}_{\textrm{fib}} = 3000$) with various linking dynamics. Each curve represents the average value computed over $10$ simulations conducted with the same set of parameters, with shading indicating the standard deviation. The time-constant of this growth is indicated with a black circle and the limit between unorganized and curved or aligned states is drawn with a dashed line.}
    \label{fig_time_Almean}
\end{figure}

Figure~\ref{fig_time_Almean} displays the temporal evolution of the quantifier $\textrm{Al}_{\textrm{mean}}$ for dense systems with various values of $\nu_{\textrm{link}}$ and $\chi_{\textrm{link}}$.

Our main observation is that, for all parameters, the alignment indicator follows an inverted exponential growth, that is a quick initial growth followed by a slow convergence towards an asymptotic value. We computed the time-constant $\tau_{\textrm{Al}}$ of this growth, that is the time needed to reach $63\%$ of the asymptotic value, and plotted it on the corresponding curve with a black circle. It can be seen that, for a given value of $\nu_{\textrm{link}}$, the shorter the time-constant, the higher the equilibrium value of the alignment indicator (compare the curves inside each panel). By comparing the panels from left to right, we see that the faster the remodeling of the network, the faster the convergence of the system towards its equilibrium value. Moreover, by comparing the extreme cases $\chi_{\textrm{link}} = 0.1$ (blue curve) with $\chi_{\textrm{link}} = 0.9$ (pink curves) of panels A and C, we see that the dependence of the reorganization time $\tau_{\textrm{Al}}$ on the equilibrium linked fiber fraction is not trivial. Indeed, fast remodeling networks (panel C) seem to reorganize faster when the equilibrium linked fiber fraction is large (pink curve) than low (blue curve), while the reverse is observed for slow remodeling networks (panel A). Altogether, these results suggest that for each network dynamics, there exists a most efficient range of equilibrium linked fiber fraction allowing for quicker convergence to equilibrium.

\begin{figure}
    \centering
    \includegraphics[width=\textwidth]{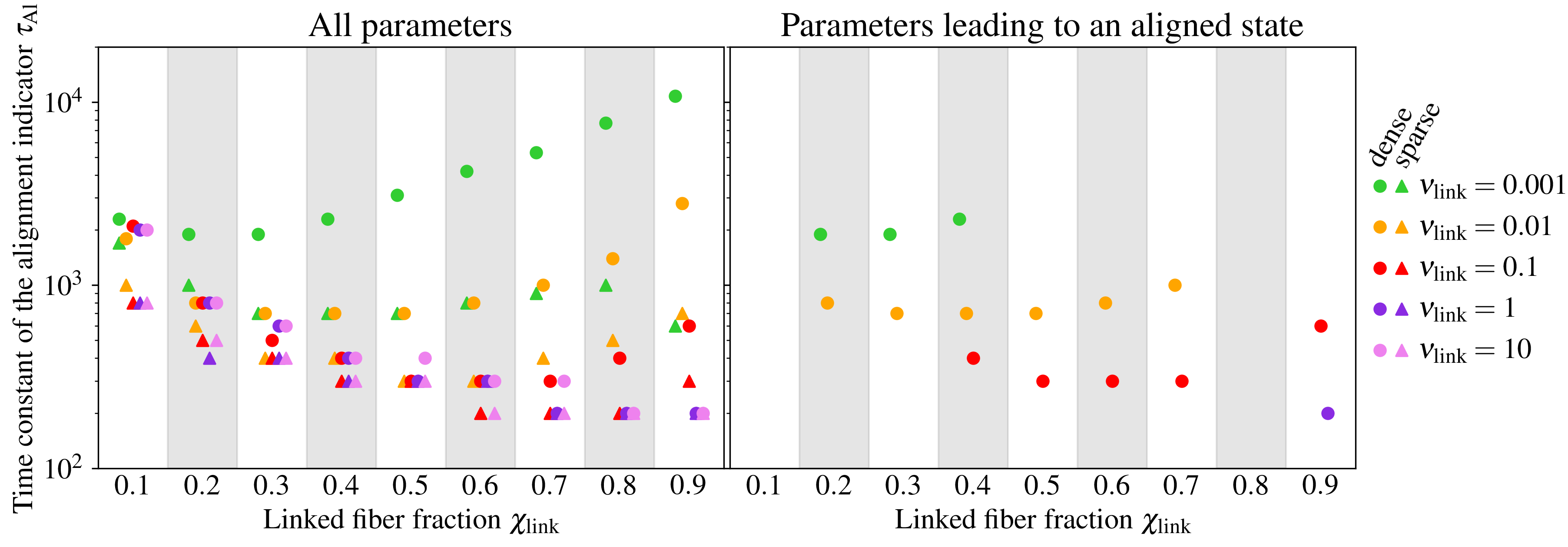} 
    \caption{Time-constant of the alignment indicator $\tau_{\textrm{Al}}$ according to the equilibrium linked fiber fraction $\chi_{\textrm{link}}$, with color depending on the remodeling speed $\nu_{\textrm{link}}$ and marker depending on the fiber density. The displayed values correspond to the average over $10$ simulations conducted with the same set of parameters. \textbf{Left}~: Results for all tested sets of parameters, whatever the outcomes of the simulations. \textbf{Right}~: Results for all tested sets of parameters which lead, on average, to an aligned state.}
    \label{fig_corrTaual}
\end{figure}

To explore in more details the dependence between the convergence speed and the parameters of the networks, in Figure~\ref{fig_corrTaual} we plot $\tau_{\textrm{Al}}$ as a function of $\chi_{\textrm{link}}$, for different values of $\nu_{\textrm{link}}$. The left panel contains all the simulations while the right panel only shows the results for the sets of parameters leading, on average over $10$ simulations, to an aligned equilibrium state (i.e. $\textrm{Al}_{\textrm{mean}} > 0.95$).

We can first see on the left panel of Figure~\ref{fig_corrTaual} that $\tau_{\textrm{Al}}$ decreases when $\nu_{\textrm{link}}$ increases according to a non-linear relationship which saturates for $\nu_{\textrm{link}} \ge 0.1$ (compare the different color points). These results show that fast remodeling networks relax faster to their steady-states than slow-dynamical networks. Moreover, sparse systems organize quicker than dense systems at low linking dynamics ($\nu_{\textrm{link}} \le 0.01$, compare the dot and triangle markers for the green and yellow populations), while there is no difference between dense and sparse systems for fast remodeling networks ($\nu_{\textrm{link}} \ge 0.1$ where dot and triangle markers are superimposed).

For each value of $\nu_{\textrm{link}}$, there is a most efficient range of equilibrium linked fiber fraction $\chi_{\textrm{link}}$ allowing for a lower value of $\tau_{\textrm{Al}}$ and so a quicker convergence to equilibrium. For slow remodeling networks ($\nu_{\textrm{link}} = 0.001$, green markers) this range lays between $\chi_{\textrm{link}} = 0.2$ and $\chi_{\textrm{link}} = 0.3$, because systems with too much crosslinks will undergo stiffening and take longer to relax, but systems with too few crosslinks will have difficulty to align themselves. As one can observe, the range of $\chi_{\textrm{link}}$ allowing the fastest convergence to equilibrium shifts towards $1$ as the network remodeling speed $\nu_{\textrm{link}}$ increases. As the network remodeling increases, a greater number of crosslinks will then promote a quicker alignment.

When looking only at parameter sets which, on average, lead to aligned equilibrium states (right panel of Figure~\ref{fig_corrTaual}), we can see that these parameter sets cover all remodeling dynamics and correspond to the range of equilibrium linked fiber fraction leading to fastest convergence for each remodeling speed. We conclude that the most efficient systems (which organize the fastest) are also those that align most.

\paragraph*{Ethics :} This article does not present research with ethical considerations.

\paragraph*{Data Access :} The code used to perform numerical simulations of our model can be found on \href{https://github.com/chassonnery/3D\_DynamicalFiberNetwork/tree/e76c323089d1041d30aa68756447f681bd38c295/Code}{GitHub}. Supplementary data are also available \href{https://github.com/chassonnery/3D\_DynamicalFiberNetwork/tree/e76c323089d1041d30aa68756447f681bd38c295/Supplementary\%20data\%20-\%20Movies}{online}.

\paragraph*{Competing interests :} The authors declare no competing interests.

\paragraph*{Funding :} This study has been partially supported through the grant EUR CARe N°ANR-18-EURE-0003 in the framework of the Programme des Investissements d'Avenir, by Sorbonne Alliance University with an Emergence project MATHREGEN, grant number S29-05Z101 and by Agence Nationale de la Recherche (ANR) under the project grant number ANR-22-CE45-0024-01.

\paragraph*{Acknowledgment :} P. Degond holds a visiting professor association with the Department of Mathematics, Imperial College London, UK.

\bibliographystyle{plain}
\bibliography{bibliography}

\end{document}